\documentclass[aps,prl,twocolumn,floatfix,superscriptaddress,amsmath,amssymb]{revtex4-1}

\usepackage{graphicx}
\usepackage{bm}
\usepackage{hyperref}
\usepackage[mathlines]{lineno}
\usepackage{braket}
\usepackage{bbm}
\usepackage{verbatim}
\usepackage{ulem}
\usepackage{standalone}
\usepackage{comment}
\usepackage[height=8.85in,width=6.45in]{geometry}
\usepackage{slashed}
\usepackage{braket}
\usepackage{tikz}
\usepackage{tikz-cd}
\usepackage{times}
\usepackage{courier}
\usepackage{mdframed}
\usepackage{datetime}
\usepackage{dsfont}
\usepackage{multirow}
\usepackage{cancel}
\usepackage{caption}
\usepackage{subcaption}
\usepackage[compat=1.1.0]{tikz-feynman}
\usepackage{pifont}
\usepackage{array} 

\newcommand{\parR}[1]{\textbf{\textit{#1}}---}

\newcommand{\Z}{\mathbb{Z}}

\renewcommand\[{\begin{equation}}
\renewcommand\]{\end{equation}}

\def\ie{\begin{equation}\begin{aligned}}
\def\fe{\end{aligned}\end{equation}}

\begin{document}

\title{Duality viewpoint of noninvertible symmetry protected topological phases}	

\newcommand{\ugent}[0]{Department of Physics and Astronomy, University of Ghent, 9000 Ghent, Belgium}
\newcommand{\penn}[0]{Department of Physics, The Pennsylvania State University, University Park, Pennsylvania 16802, USA}
\newcommand{\ipmu}[0]{Kavli Institute for the Physics and Mathematics of the Universe, University of Tokyo, 5-1-5 Kashiwanoha, Kashiwa, Chiba, 277-8583, Japan}
\newcommand{\utokyo}[0]{Department of Physics, Graduate School of Science, University of Tokyo, Tokyo 113-0033, Japan}
\newcommand{\tsqs}[0]{Trans-Scale Quantum Science Institute, University of Tokyo, Tokyo 113-0033, Japan }
\newcommand{\cqm}[0]{Center for Quantum Mathematics at IMADA, Southern Denmark University, Campusvej 55, 5230 Odense, Denmark}
\newcommand{\nbi}[0]{Niels Bohr International Academy, Niels Bohr Institute, University of Copenhagen, Denmark}

\author{Weiguang Cao}
\affiliation{\cqm}
\affiliation{\nbi}

\author{Masahito Yamazaki}
\affiliation{\ipmu}
\affiliation{\utokyo}
\affiliation{\tsqs}

\author{Linhao Li}
\email{linhaoli601@gmail.com}
\thanks{Corresponding author.}
\affiliation{\ugent}
\affiliation{\penn}
\date{\today}

\begin{abstract}
Recent advancements in generalized symmetries have drawn significant attention to gapped phases of matter exhibiting novel symmetries, such as noninvertible symmetries. By leveraging the duality transformations, the classification and construction of gapped phases with noninvertible symmetry can be mapped to those involving conventional group symmetries. We demonstrate this approach by classifying symmetry-protected-topological phases with a broad class of noninvertible symmetries in arbitrary spacetime dimensions. Our results reveal new classifications that extend beyond those based on group symmetries. Additionally, we construct lattice models in $(1+1)D$ and $(2+1)D$ that realize these new phases and explore their anomalous interfaces.
\end{abstract}
\maketitle

\parR{Introduction.} {Symmetries play a central role in understanding phases of matter. Recent developments in noninvertible symmetries—described by fusion categories and characterized by operations that lack an inverse—have extended the traditional group-based classification of phases to a more general, category-theoretic framework. These noninvertible symmetries appear ubiquitously in physics, with significant applications in particle physics
~\cite{Choi:2022jqy,Cordova:2022ieu,PhysRevLett.128.111601,PhysRevD.105.125016,PhysRevLett.130.131602,Kaidi:2022uux,Bhardwaj:2022yxj,Bhardwaj:2022lsg,Bhardwaj:2022kot,Bartsch:2022mpm}, quantum lattice models~\cite{
Aasen:2016dop,Aasen:2020jwb, Chen:2023qst, Lootens_2023, Lootens_2024, Eck_Fendley_1,PhysRevB.107.125158, Li:2023knf,Li_2023, Cao:2023doz,Seiberg:2023cdc, Sinha:2023hum, Seiberg:2024gek,Yan:2024eqz, Seifnashri:2023dpa, Tavares:2024vtu, Cao:2024qjj,Lu:2024ytl,Pace:2024tgk, Pace:2024acq, Pace:2024oys, Okada:2024qmk, Seo:2024its,Cao:2025qnc}. 
and novel phases which benefit the design of quantum devices~\cite{ Warman:2024lir,Lootens:2024gfp,Bhardwaj:2023idu,Bhardwaj:2023fca,Bhardwaj:2023bbf,Bhardwaj:2024qrf,Bhardwaj:2024wlr,Bhardwaj:2024kvy,Bhardwaj:2024qiv,Chatterjee:2024ych,Warman:2024lir,Bhardwaj:2025piv}.}

{Among these novel phases, symmetry-protected-topological (SPT) phases serve as important theoretical bridges between different disciplines of physics. SPT was initially defined as gapped phases with a unique ground-state with ordinary group symmetries~\cite{PhysRevB.80.155131,PhysRevLett.95.226801,PhysRevLett.95.146802,Pollmann:2009ryx,Pollmann:2009mhk,Chen:2010zpc,PhysRevB.84.235128,Chen:2011bcp,Chen:2011pg}.
Noninvertible symmetries further enrich the classification of SPT phases and enable new lattice realizations ~\cite{Choi:2024rjm,Seifnashri:2024dsd,Jia:2024bng,Li:2024fhy,Jia:2024zdp,Jia:2024wnu,Inamura:2024jke,Meng:2024nxx,Li:2024gwx}.
These new phases offer promising resource states for measurement-based quantum computation~\cite{Fechisin:2023dkj}, motivating the development of systematic tools to distinguish and classify such noninvertible symmetry-protected-topological (NISPT) phases.}

{A powerful and general method for distinguishing SPT phases with group symmetries~\cite{levin2012braiding,wen2014symmetry,PhysRevLett.112.141602,PhysRevB.90.035451,barkeshli2019symmetry} utilizes duality such as gauging the group symmetries.
For example, the nontrivial $\mathbb Z_2\times\mathbb Z_2$ SPT phase can be studied through $\mathbb Z_2\times\mathbb Z_2$ spontaneous-symmetry-breaking (SSB) phase via the Kennedy–Tasaki (KT) duality~\cite{Kennedy:1992tke,PhysRevB.45.304,oshikawa1992hidden,Li_2023}, also known as twist gauging. More recently, KT duality also relates three distinct $(1+1)D$ Rep($D_8$) NISPT phases~\cite{Seifnashri:2024dsd} to different SSB patterns of the dual $(\mathbb{Z}^{(0)}_2)^3$ symmetry. However, a complete  classification of NISPT phases, together with their higher-dimensional lattice realization, is still lacking.}

\begin{figure}[t]
    \centering
\includegraphics[width=1.0\linewidth]{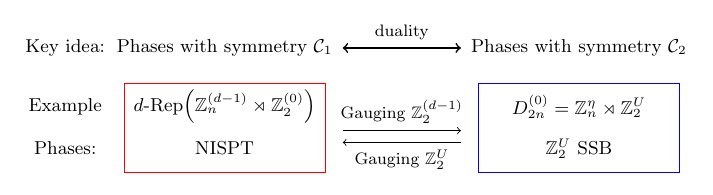}
    \caption{{The duality method to study NISPT phases from group SSB phases with examples in the boxes.}
    }
    \label{fig:dualityfigure}
\end{figure} 

{In this letter, we take a first step toward a systematic classification by extending the duality method to study NISPT phases in general dimensions. The duality between  symmetries $\mathcal C_1$ and $\mathcal C_2$ establishes a one-to-one correspondence between gapped phases on two sides, thereby enabling SPTs with symmetry $\mathcal C_1$ to be constructed from gapped phases of $\mathcal C_2$.  As illustrated in Fig.~\ref{fig:dualityfigure}, we focus on one of the simplest yet most intriguing cases: gauging $\mathbb Z_2^{(0)}$ symmetry in the systems with symmetry $\mathcal{C}_2$ as $D_{2n}^{(0)}$, the dihedral group of order $2n$.} 
In $(d+1)D$, the dual symmetry $\mathcal C_1$ is described by a fusion $d$-category $d$-Rep$(\mathbb{Z}^{(d-1)}_n\rtimes \mathbb{Z}^{(0)}_2)$, which is among the {most useful} families of noninvertible symmetries.
For $d=1,2$, this symmetry has been extensively studied in lattice models 
~\cite{Bhardwaj:2023idu,Bhardwaj:2023fca,Bhardwaj:2023bbf,Bhardwaj:2024qrf,Bhardwaj:2024wlr,Bhardwaj:2024kvy,Bhardwaj:2024qiv,Chatterjee:2024ych,Warman:2024lir,Bhardwaj:2025piv,Seifnashri:2024dsd,Choi:2024rjm}, and {quantum simulation proposals with } Rydberg atom arrays~\cite{ Warman:2024lir}. For $d=3$, it has recently attracted a lot of attention in field theory and string theory~\cite{Kaidi:2024wio}, with plenty of applications in particle phenomenology~\cite{Kobayashi:2024cvp,Kobayashi:2025znw,Liang:2025dkm,Kobayashi:2025cwx}. 
{Despite this broad interest}, 
the corresponding gapped phases, particularly NISPT phases and the bulk-edge correspondence, remain largely unexplored.

To fill the gap, we establish a complete classification for NISPT phases with symmetries $\mathcal C_1$, 
by leveraging traditional group-theoretic classifications of SSB and SPT phases with $\mathcal C_2$ group symmetries. It is known that NISPTs are classified by the fiber functors of 
noninvertible symmetries~\cite{Thorngren:2019iar}, extending the  cohomological classification of group SPTs~\cite{Chen:2011pg}. Our classification of NISPTs has  different branches originated from different dual SSB patterns, providing a concrete and physical understanding of the fiber functors.

Moreover, we turn to explicit lattice realizations in $(1+1)D$ and $(2+1)D$. Employing the duality method, we construct new stabilizer models realizing these NISPTs {in  tensor-product Hibert space}. {These models further allow the study of  interfaces between distinct NISPTs and anomaly-inflow with these noninvertible symmetries, which for the first time demonstrates the bulk-edge correspondence of $(2+1)D$ NISPTs.}

\parR{Classification of NISPTs dual to $\mathbb Z_2$-SSB phases with $D_{2n}$ 0-form symmetry.}We start with models possessing non-abelian 0-form symmetry 
$D_{2n}=\mathbb Z_n^{\eta}\rtimes \mathbb Z_2^{U}$, generated by $\mathbb Z_n$ operator $\eta$ and $\mathbb Z_2$ operator $U$, in $(d+1)$ dimensions{~\footnote{The superscript $(0)$ for 0-form symmetry is omitted when unambiguous.}}. 
We gauge the non-normal $\mathbb Z_2^U$ subgroup and the dual models exhibit non-invertible symmetries. 

For $d=1$, gauging $\mathbb Z_2^{U}$ symmetry, implemented by the Kramers-Wannier (KW) duality~\cite{Kramers:1941kn,Frohlich_2004}, leads to Rep$(D_{2n})$ symmetry. When $d=2$, the dual symmetry is $2$-Rep$(\mathbb{Z}^{(1)}_n\rtimes \mathbb{Z}^{(0)}_2)$~\cite{Decoppet:2024htz,Choi:2024rjm,Bhardwaj:2022maz}. We conjecture the dual noninvertible symmetry to be $d$-Rep$(\mathbb{Z}^{(d-1)}_n\rtimes \mathbb{Z}^{(0)}_2)$ in general dimensions. The symmetry operators correspond to the irreducible (higher) representations of the (higher) group $\mathbb Z_n^{(d-1)}\rtimes \mathbb Z_2^{(0)}$.  In the appendix, we derive the symmetry operators and their fusion algebra by gauging $\mathbb Z_2^U$ on the lattice. {Notably, the $\Z^U_2$-even combination of $\Z_n$ operators $\eta^l+\eta^{-l}$ can be made gauge invariant.
After gauging, this construction yields the noninvertible symmetry operators $\mathsf{W}_l$ for $l\neq n/2$, while $\eta^{n/2}$ gives rise to a $\mathbb{Z}_2$ symmetry operator (for $n=0$~mod~$2$).
The noninvertible fusion rule
\begin{equation}
\mathsf{W}_l\times \mathsf{W}_k=\mathsf{W}_{l+k}+\mathsf{W}_{l-k}, \quad l,k\in \Z_n
\end{equation}
directly follows from the fusion rule of these $\Z^U_2$-even combinations.  In addition, gauging also gives rise to a quantum $\mathbb{Z}_2^{(d-1)}$ symmetry.}

The gapped phases with dual $d$-Rep$(\mathbb{Z}^{(d-1)}_n\rtimes \mathbb{Z}^{(0)}_2)$ symmetry can be studied by the corresponding gapped phases with $D_{2n}$ symmetry. In particular, NISPTs correspond to SSB phases where two-fold degenerate ground states break the $\mathbb Z_2^U$ subgroup. 
These SSB phases are firstly distinguished by conjugacy classes of unbroken-subgroup $\Gamma$
\[
[\Gamma
]=\{g\Gamma
g^{-1}|\Gamma\subset D_{2n}, g\in D_{2n}\}.
\]
Specifically, the unbroken-subgroup $\Gamma$ of $\Z^U_2$-SSB phases are all normal subgroups (i.e., $[\Gamma
]=\{\Gamma
\}$):
\begin{enumerate}
\item  If $n$ is odd,  $\Gamma=\mathbb{Z}_n^\eta$ .
\item If $n$ is even, there are two choices. Besides $\mathbb{Z}_n^{\eta}$, the other is $\mathbb Z_{n/2}^{\eta^2}\rtimes\mathbb Z_2^{\eta U}=\{\eta^2|\eta^n=1\}\rtimes \{1,\eta U\}$. 
\end{enumerate}
Moreover, $\Z^U_2$-SSB phases are further distinguished by 
 the conjugacy classes of attached $\Gamma$-SPT phases $\Omega^{\Gamma}$  \cite{Aksoy:2025rmg}:
\[\label{eq:conjugacy classes}
[\Omega^\Gamma]=\{g\Omega^\Gamma |\Omega^\Gamma\in H^{d+1}(\Gamma,U(1)), g\in D_{2n}\}.
\]
Here $g\Omega^\Gamma $ denotes the $\Gamma$-SPT obtained by applying the $g$-transformation to a system in the phase $\Omega^\Gamma$.

{
After the duality transformation, different classes of $\Z^U_2$-SSB phases are mapped one-to-one to NISPT phases. 
Our classification, especially for phases in different branches of unbroken-subgroups, is not described by a direct product of abelian groups and is therefore beyond the ordinary group-cohomology classification. This is consistent with the non-stacking feature of NISPTs~\cite{Seifnashri:2024dsd}.
We leave the intriguing mathematical structure behind this classification for future work and only keep track of the numbers in different branches, which are exemplified in Table.~\ref{tab:SPTclassification} for $d=1,2,3$. Further details are provided in the appendix.
}

\begin{table}[!tbp]
    \centering
    \begin{tabular}{|>{\centering\arraybackslash}p{2.0cm}|>{\centering\arraybackslash}p{1.0cm}|>{\centering\arraybackslash}p{2.0cm}|>{\centering\arraybackslash}p{1.0cm}|}
    \hline
      Classification  & $d=1$ & $d=2$  & $d=3$ \\
    \hline
         $n=1$ mod 2  & $1$ & $n$ & $1$ \\
    \hline
         $n=2$ mod 4  & $1+1$ & $n+n$ & $1+1$ \\
    \hline
         $n=0$ mod 4  & $1+2$ & $n+3n/2$ & $1+3$ \\
    \hline
    \end{tabular}
    \caption{{Classification} of $d$-Rep$(\mathbb{Z}^{(d-1)}_n\rtimes \mathbb{Z}^{(0)}_2)$ SPTs. {We show only the number of different classes of NISPT in each branch.}}
    \label{tab:SPTclassification}
\end{table}

\parR{Rep$(D_{2n})$ symmetry in spin chain.}Here we construct Rep$(D_{2n})$ symmetry in the $(1+1)D$ spin chain with $L$ sites and a tensor product Hilbert space $(\mathbb C^2)^L$, in preparation for discussion of Rep$(D_{2n})$-SPT for next section.
Starting from $D_{2n}$ symmetry generated by 
\begin{equation}
 \eta=\prod_{j=1}^L\exp\left(\frac{\pi i}{n}(1-\sigma^z_j)\right), \quad U=\prod_{j=1}^{L}\sigma^x_j,
\end{equation}
we perform the KW duality transformation
\begin{equation}\label{eq:kw1dlattice}
\mathcal{D}: \sigma^x_{j}\to \sigma^z_{j-1}\sigma^z_{j},\quad \sigma^z_{j}\sigma^z_{j+1}\to \sigma^x_{j},
\end{equation}
and obtain the dual  Rep$(D_{2n})$ symmetry.  The objects of  Rep$(D_{2n})$ are irreducible representations (irreps) of $D_{2n}$ and their fusion algebra follows from the tensor product of these irreps. When $n$ is even, $D_{2n}$ has four $1d$ irreps and $(n/2-1)$ $2d$ irreps. The dual Rep$(D_{2n})$ symmetry is generated by
\begin{align}
    \begin{split}
        &U^e=\prod_{k=1}^{L/2}\sigma^x_{2k},\quad U^o=\prod_{k=1}^{L/2}\sigma^x_{2k-1},\\
        &\mathsf{W}_{l}=\mathcal{D}^{\dagger}(\eta^l+\eta^{-l})\mathcal{D},\quad l=1,...,n/2-1,
    \end{split}
\end{align}
where $U^e,U^o$ generate the invertible $\mathbb Z_2\times \mathbb Z_2$ subsymmetry and $\mathsf{W}_{l}$, constructed by sandwich method~\cite{Cao:2025qnc}, generates the noninvertible part. 
When $n$ is odd, $D_{2n}$ has two $1d$ irrep and $(n-1)/2$ $2d$ irreps, and the dual symmetry is generated by
\begin{equation}
    U=\prod_{j=1}^L\sigma^x_{j},\quad \mathsf{W}_l=\mathcal{D}^{\dagger}(\eta^{l}+\eta^{-l})\mathcal{D},\, l=1,...,(n-1)/2.\nonumber
\end{equation}

{\parR{Lattice realization of Rep($D_{2n}$) SPTs.} In this section, we give a complete construction of Rep$(D_{2n})$-SPT phases using the KW duality. We start from the $\Z^U_2$-SSB side and}
the SPT phases of unbroken-subgroup are classified by 
\begin{align}
    \begin{split}  
H^{2}(\mathbb Z_{n},U(1))&=\mathbb Z_1,\\ H^{2}(\mathbb Z_{n/2}\rtimes \mathbb Z_2,U(1))&=\mathbb Z_{\text{gcd}(n/2,2)},
   \end{split}
\end{align}
{and we show in the appendix that all such phases are invariant under the $D_{2n}$ transformation.} Thus we obtain the classification of Rep($D_{2n}$)-SPT phases in Table.~\ref{tab:SPTclassification} for $d=1$.
These SPT phases can be explicitly constructed from  $\mathbb Z_2^U$-SSB models by performing KW duality \eqref{eq:kw1dlattice}. 

The simplest case is the phase in $H^{2}(\mathbb Z_{n}, U(1))=\mathbb Z_1$ where the ground states preserve the $\mathbb Z_n^{\eta}$ symmetry, for example, realized by the ordered phase of Ising model
$-\sum_{j}\sigma^z_{j}\sigma^z_{j+1}$.
The dual Rep$(D_{2n})$ SPT phase is the product state $H_{\text{prod}}=-\sum_{j}\sigma^x_j$.

The next case is the phase in $H^{2}(\mathbb Z_{n/2}\rtimes \mathbb Z_2,U(1))$ where the ground state preserves the symmetry operators $\eta^2$ and $\eta U$.

{Let's first consider the simplest example with $n=4$ involving two distinct SPTs. Assuming $L=0$ mod $2$, we group $2$ qubits in one unit-cell. This type of $\mathbb Z_2^U$-SSB phase is realized by the stabilizer Hamiltonian
\begin{widetext}
    \begin{equation}\label{eq:evenoddphased8}
    H=-\sum_{k=1}^{L/2}\sigma_{2k-1}^z\sigma_{2k}^{z}-\sum_{k=1}^{L/2}(\sigma^y_{2k-1}\sigma^x_{2k}\sigma^x_{2k+1}\sigma^y_{2k+2}+\sigma^x_{2k-1}\sigma^y_{2k}\sigma^y_{2k+1}\sigma^x_{2k+2})
\end{equation}
\end{widetext}
where the first term aligns the spins within the same unit-cell in the ferromagnetic configuration, reducing the ground state degeneracy (GSD) to $2^{L/2}$, and the second term introduces $D_{8}$ symmetric interaction to further reduce GSD to 2. To check this statement, we treat two  ferromagnetic  spins in each unit cell as single spin in the low energy:
    \begin{equation}\label{eq: subspace}
   |\tilde{\uparrow}\rangle_k=|\uparrow\uparrow\rangle_{2k-1,2k},\quad |\tilde{\downarrow}\rangle_k=|\downarrow\downarrow\rangle_{2k-1,2k},
    \end{equation}
and denote $\tilde{\sigma}^x_k,\tilde{\sigma}^z_k$ for the Pauli operators acting on the $k$-th unit cell. We sketch this process in Fig.~\ref{fig:effectivefigure} and list the dictionary between the symmetry operators in the original lattice and in the effective theory. 
\begin{figure}[t]
    \centering
\includegraphics[width=0.9\linewidth]{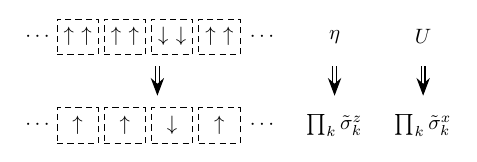}
\caption{{Effective process (left) and the dictionary of symmetry operators in UV and IR (right).}
    }  \label{fig:effectivefigure}
\end{figure}
The SSB Hamiltonian \eqref{eq:evenoddphased8} effectively becomes
\begin{equation}
    \tilde{H}=-2\sum_{k=1}^{L/2}\tilde{\sigma}^y_k\tilde{\sigma}^y_{k+1},
\end{equation}
which preserves $\mathbb Z_2^{\eta^2}\times \mathbb Z_2^{\eta U}$ and spontaneously breaks the $\mathbb Z_2^U$. After KW transformation, the SSB phase \eqref{eq:evenoddphased8} becomes the Rep$(D_{8})$ \textit{even} SPT phase~\footnote{Our convention is different from the one in~\cite{Seifnashri:2024dsd} by a unitary transformation to every SPT phase.}
\begin{equation}\label{eq:evenHAL}
     H_{\text{even}}=-\sum_{k=1}^{L/2}\sigma^x_{2k-1}-\sum_{k=1}^{L/2}\sigma^z_{2k-2}\sigma^x_{2k}\sigma^z_{2k+2}(1+\sigma^x_{2k-1}\sigma^x_{2k+1}),
\end{equation}
and the \textit{odd} SPT is obtained by one-site translation on \eqref{eq:evenHAL}.

Now we construct the models with general even $n$ using the same idea. 
Assuming $L=0$ mod $n/2$, we group $n/2$ qubits in one unit-cell. The $\mathbb Z_2^U$ SSB phase is realized by the stabilizer Hamiltonian
\begin{equation}\label{eq:evenoddphase}
    H=H_{\text{1}}+H_{\text{2}},
\end{equation}
where the first term
\begin{equation}
    H_{\text{1}}=-\sum_{k=1}^{2L/n}\left(\sum_{j=1}^{n/2-1}\sigma^z_{nk/2-j}\sigma^z_{nk/2-(j-1)}\right),
\end{equation}
aligns the spins within the same unit-cell in the ferromagnetic configuration, reducing the ground state degeneracy (GSD) to $2^{2L/n}$. 
The second term
\begin{equation}
    H_2=\sum_{l=0}^{ n/2 -1}\eta^{-l}H_2^{(0)}\eta^{l},
\end{equation}
introduces $D_{2n}$ symmetric interaction to further reduce GSD to 2, where 
\begin{equation}\label{eq:unsymint}
    H_2^{(0)}=\sum_{k=1}^{2L/n-1}\sigma^y_{\frac{n(k-1)}{2}+1}\left(\prod_{j=\frac{n(k-1)}{2}+2}^{\frac{n(k+1)}{2}-1}\sigma^x_{j}\right)\sigma^y_{\frac{n(k+1)}{2}},
\end{equation}
acts on the ground states effectively as $\sum_{k}\tilde{\sigma}^y_k\tilde{\sigma}^y_{k+1}$. 
For $n=2$ mod 4, the classification is $\mathbb Z_1$ and \eqref{eq:evenoddphase} is the only phase. For $n=0$ mod 4, the classification is $\mathbb Z_2$, and the other phase is obtained by one-site translation on \eqref{eq:evenoddphase}, with 
nontrivial edge modes shown in the appendix.

After KW transformation, the SSB Hamiltonian \eqref{eq:evenoddphase} becomes the Rep$(D_{2n})$ SPT Hamiltonian
\begin{equation}\label{eq:evennispt}
    \tilde{H}=-\sum_{k=1}^{2L/n}\left(\sum_{j=1}^{ n/2-1}\sigma^x_{nk/2-j}\right)+\tilde{H}_2
\end{equation}
where 
$\tilde{H}_2=\sum_{l=0}^{n/2-1}\tilde{H}_2^{(l)}$ is the Rep$(D_{2n})$ symmetric interaction with $\tilde{H}_2^{(0)}\mathsf{W}_{l}=\mathsf{W}_{l}\tilde{H}_2^{(l)}$ and 
\begin{equation}\label{eq:repd2nspt}
    \tilde{H}_2^{(0)}=-\sum_{k=1}^{2L/n}\sigma^z_{n(k-1)/2}\left(\prod_{j=n(k-1)/2+1}^{n(k+1)/2-1}\sigma^x_{j}\right)\sigma^z_{n(k+1)/2},\nonumber
\end{equation}
is the KW dual of \eqref{eq:unsymint}. For $n=0$ mod 4, we obtain the other SPT by one-site translation.  
}

\parR{Projective representation at the interface.}  One signature of SPT phases is the edge modes, arising from the projective representation of the symmetries at the interfaces between two distinct SPTs. Consider a closed chain with $2L$ sites and put the Rep($D_{2n}$) SPT on the first $L$ sites and the $2L$-th site, and the product state on the remaining sites. As shown in the appendix, the action of symmetry operators on ground states of the Hamiltonian\eqref{eq:evennispt} factorize at the interfaces when $n=0$ mod 4
\begin{align}
    \begin{split}
        &U^o\ket{\text{GS}}=\ket{\text{GS}},\\        &U^{e}\ket{\text{GS}}=\sigma^y_{L}\sigma^y_{2L}\ket{\text{GS}}:=U^{e,(\text{L})}U^{e,(\text{R})}\ket{\text{GS}},\\
        &\mathsf{W}_{n/4}\ket{\text{GS}}=(\sigma^x_{L}\sigma^z_{2L}+\sigma^z_{L}\sigma^x_{2L})\ket{\text{GS}}\\
        &:=(\mathsf{W}_{n/4}^{(\text{L}),1}\mathsf{W}_{n/4}^{(\text{R}),1}+\mathsf{W}_{n/4}^{(\text{L}),2}\mathsf{W}_{n/4}^{(\text{R}),2})\ket{\text{GS}},
    \end{split}
\end{align}
and generate  projective algebra at each interface
\begin{align}
    \begin{split}
    &\mathsf{W}_{n/4}^{(\text{L}),k}U^{e,(\text{L})}=-U^{e,(\text{L})}\mathsf{W}_{n/4}^{(\text{L}),k}, \\
    &\mathsf{W}_{n/4}^{(\text{R}),k}U^{e,(\text{R})}=-U^{e,(\text{R})}\mathsf{W}_{n/4}^{(\text{R}),k}.
  \end{split}
\end{align}
{Due to this anticommutation relation, such an SPT is referred to as the \textit{even} SPT. The other SPT after one-site translation instead exhibits an anticommutation relation between $U^o$ and $\mathsf{W}_{n/4}$ at the interface and is denoted the \textit{odd} SPT.}

\parR{Noninvertible symmetry in $(2+1)D$.} The duality method is also powerful in $(2+1)D$ to construct fusion-2 category symmetries on the lattice. Our main example is $D_8$ symmetry  and its dual noninvertible symmetry described by 2-Rep$(\mathbb Z_4^{(1)}\rtimes \mathbb Z_2^{(0)})$~\cite{Bhardwaj:2022maz,Decoppet:2024htz}.
\begin{figure}[t]
    \centering
    \includegraphics[width=0.6\linewidth]{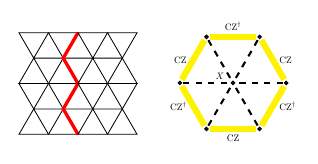}
    \caption{
   {Left: triangular lattice and interface (red) between different NISPTs. Right:  seven-body interaction term in \eqref{eq:2Dungauged Hal}.}}\label{fig: interface}
\end{figure}

Consider a triangular lattice $\Lambda$ shown in Fig.~\ref{fig: interface}. 
On each vertex, we assign a qubit $\ket{\sigma}_j, \sigma\in\mathbb Z_2$ and a $\mathbb Z_4$ qudit $\ket{s}_j, s\in\mathbb Z_4$. The clock and shift operators acting on a vertex $j$ are
\begin{equation}
Z_j=\sum_{s=0}^{3}e^{\frac{\pi i}{2}s}\ket{s}\bra{s}_j, \ X_j=\sum_{s=0}^{3}\ket{s+1}\bra{s}_j.
\end{equation}
The $D_8$ symmetry is generated by
\begin{equation}\label{eq:lattice D8 sym}
    \eta=\prod_{j\in \Lambda} X_j,\quad  U=\prod_{j\in \Lambda} C_j \sigma^x_j, 
\end{equation}
where $C_j=\sum_{s=0}^{3}\ket{-s}\bra{s}_j$ is charge conjugation on vertex $j$.

To gauge the $\mathbb{Z}^{U}_2$
symmetry, we introduce the $\mathbb{Z}_2$ gauge field $\mu_l$ on each link $l$ and project the extended Hilbert space to the gauge-invariant sector. The Gauss law on each vertex and the gauge-flatness condition on each triangle $f$ are
\begin{equation}\label{eq:flatness}
\begin{split}
G_i=C_i\sigma^x_i\prod_{i\in l}\mu^z_l=1, \quad \prod_{l\in f}\mu^x_l=1.
 \end{split}
\end{equation}

\begin{figure}[t]
	\centering
	\includegraphics[scale=0.25]{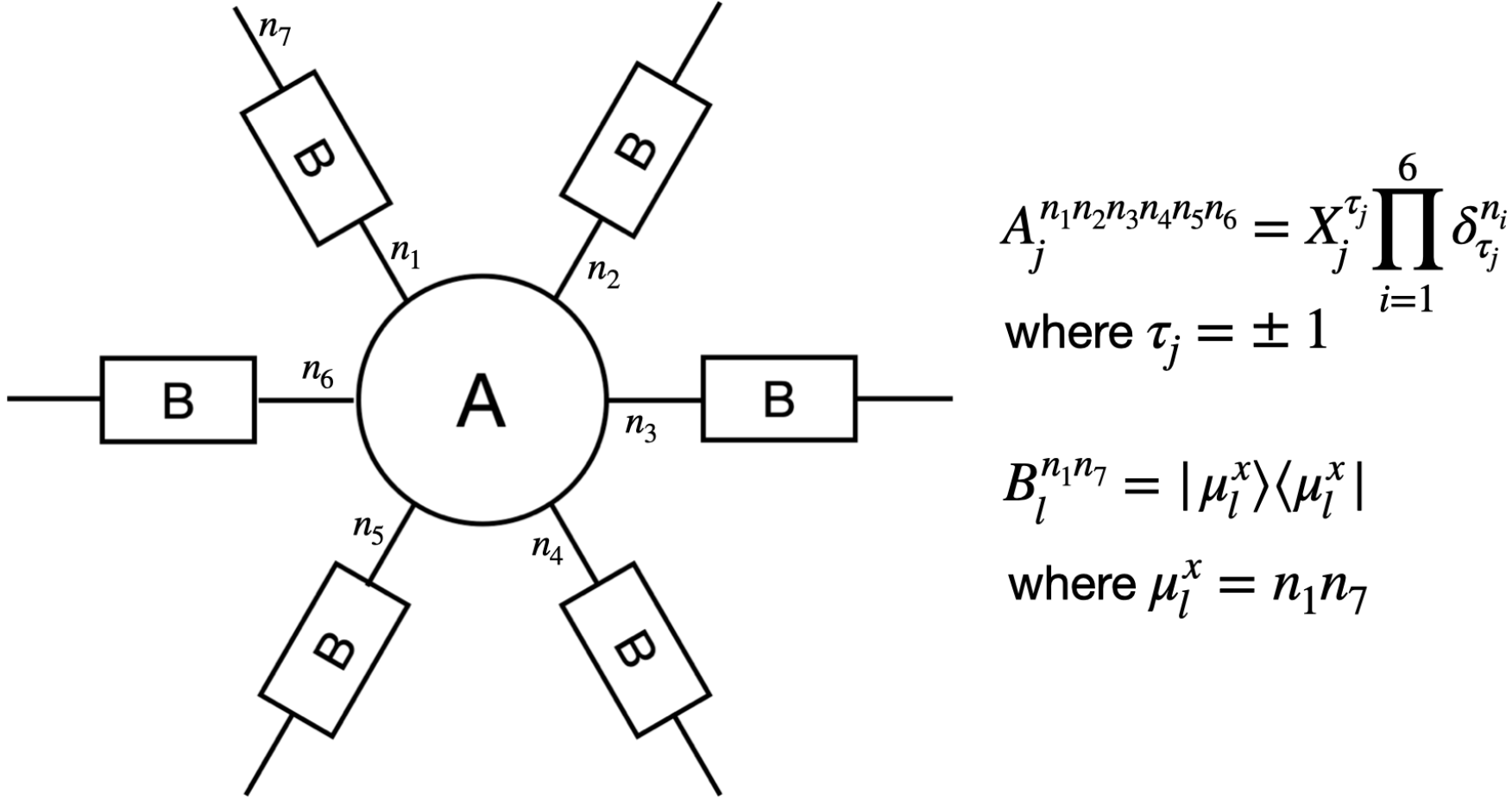}
    \captionof{figure}{ Local tensors on vertices and links. The physical bonds are in the directions perpendicular to the page. $A$-tensor acts on the vertices and $B$-tensor acts on the edges, which are connected by virtual bonds labeled by $n=\pm 1$. $\prod^6_{i=1}\delta^{n_i}_{\tau_{j}}$ is a projection operator ensuring all the  virtual bonds $n_i$ around the vertex $j$ with the same value $n_i=\tau_j=\pm 1 $.}\label{fig:local_tensor}
\end{figure}

The dual symmetry after gauging contains a quantum $\mathbb{Z}_2^{(1)}$ symmetry $W_{\gamma}=\prod_{l\in \gamma} \mu^x_l$, generated by the product of $\mu^x$ on any loop $\gamma$.
Besides, there is a $\mathbb{Z}_2^{(0)}$ symmetry operator $\eta^2$ and a 0-form  noninvertible symmetry 
\begin{equation}
\begin{split}
\mathsf{W}_1=\mathsf{C}\left(\prod_{j\in \Lambda} X^{\prod_{l\in P_{(j_0,j)}}\mu^x_l}_j+\textrm{(h.c.)}\right),
\end{split}
\end{equation}
originated from the $\mathbb Z_4$ subgroup of $D_8$.
Here $P_j$ is a path from a fixed point $j_0$ to the site $j$.  The condensation operator $\mathsf{C}=(1+W_x)(1+W_y)/4$ includes one-form symmetry $W_{\gamma}$ on non-contractible loops along $x$ and $y$ direction. The symmetry operators $W_{\gamma}$, $\eta^2$ and $\mathsf{W}_1$ generate the fusion algebra of the fusion 2-category 2-Rep$(\mathbb Z_4^{(1)}\rtimes \mathbb Z_2^{(0)})$ 
\begin{equation}
\begin{split}
&\mathsf{W}_1^2=2\mathsf{C}(1+\eta^2),\quad  \eta^2W_{\gamma}=W_{\gamma}\eta^2,\\
&\eta^2\mathsf{W}_1=\mathsf{W}_1\eta^2=W_{\gamma} \mathsf{W}_1=\mathsf{W}_1 W_{\gamma}=\mathsf{W}_1,\quad \forall \gamma\in \Lambda.\nonumber
    \end{split}
\end{equation}
We show the tensor network representation of $\mathsf{W}_1$ in Fig.~\ref{fig:local_tensor} 
with more details in the appendix.

\parR{Lattice realization of 2-Rep$(\mathbb Z_4^{(1)}\rtimes \mathbb Z_2^{(0)})$ SPTs.} We turn to concrete examples of NISPTs with 2-Rep$(\mathbb Z_4^{(1)}\rtimes \mathbb Z_2^{(0)})$ symmetry in $(2+1)D$.
From Tab.~\ref{tab:SPTclassification}, there are $4+6=10$ distinct 2-Rep$(\mathbb Z_4^{(1)}\rtimes \mathbb Z_2^{(0)})$-SPTs and we focus on two of them in the first branch in this section. To construct dual NISPTs, we first consider  $\mathbb Z_2^U$-SSB phases distinguished by stacked distinct $\mathbb{Z}^{\eta}_4$-SPT phases \footnote{This nontrivial SPT phase with $m=1$ corresponds to the generator of $\mathbb Z_2$
subgroup of $\mathbb Z_4$ classification. See the appendix for related details.}:
\begin{widetext}
\begin{equation}\label{eq:2Dungauged Hal}
\begin{split}
    H_m=-\sum_{\langle i,j\rangle\in \Lambda}\sigma^z_i\sigma^z_j-\sum_{i\in \Lambda} X_{i}\left(\prod_{(i,j,k)\! \in\bigtriangleup} \text{CZ}_{j,k}\prod_{(i,j,k)\!\in\bigtriangledown} \text{CZ}_{j,k}^{\dagger}\right)^m+ \textrm{(h.c.)}, \quad m=0,1,
\end{split}
\end{equation}
\end{widetext}
where $\bigtriangleup/\bigtriangledown$ represents the sets of all up/down triangles,  $\langle i,j\rangle$ labels the nearest neighbored pair of vertices, and $\text{CZ}_{j,k}=\sum^3_{\alpha,\beta=0}\omega^{-\alpha\beta}Z^{\alpha}_jZ^{\beta}_k$ is the controlled-Z gate.

Minimally coupling the Hamiltonian~\eqref{eq:2Dungauged Hal} with gauge field $\mu_l$, we get the gauge-invariant Hamiltonian~\cite{Chatterjee:2024ych}
\begin{equation}
\begin{split}
&H^{\text{gauged}}_m
=-\sum_{\langle i,j\rangle\in \Lambda}\sigma^z_i\mu^x_{\langle i,j\rangle}\sigma^z_j\\&-\sum_{i\in \Lambda} X_{i}\left(\prod_{(i,j,k)\in\bigtriangleup} \!\text{CZ}^{\mu}_{i,j,k}
\! \prod_{(i,j,k)\in\bigtriangledown}\!(\text{CZ}^{\mu}_{i,j,k})^{\dagger}\right)^m\!\!+\textrm{(h.c.)},\nonumber
\end{split}
\end{equation}
where $\text{CZ}^{\mu}_{i,j,k}=\sum^3_{\alpha,\beta=0}\omega^{-\alpha\beta}Z^{\alpha \mu^x_{\langle i,j\rangle}}_jZ^{\beta\mu^x_{\langle i,k\rangle}}_k$. 
Further applying the unitary transformation
\begin{equation}
  \mathcal U=\prod_{j\in \Lambda} \left(\frac{(1+\sigma^z_j)}{2}+\frac{(1-\sigma^z_j)}{2}C_j\prod_{ l\ni j}\mu^z_l\right),
\end{equation}
to simplify the Gauss law to $\sigma^x_i=1$, and projecting onto the gauge-invariant sector, we get the NISPT model 
\begin{equation}\label{eq:spt2+1}
\begin{split}
    &H^{\text{SPT}}_m
   =-\sum_{\langle i,j\rangle}\mu^x_{\langle i,j\rangle}\\&-
   \sum_i X_{i}\left(\prod_{(i,j,k)\in\bigtriangleup} \!\text{CZ}^{\mu}_{i,j,k}\prod_{(i,j,k)\in\bigtriangledown}\! (\text{CZ}^{\mu}_{i,j,k})^{\dagger}\right)^m \!\!+
   \textrm{(h.c.)},\nonumber
\end{split}
\end{equation}
defined in the constrained Hilbert space satisfying the flat gauge field condition~\eqref{eq:flatness} on each triangle~\footnote{In this constrained Hilbert space, the $\mathbb{Z}_2$ 1-form symmetry is topological.}.
The ground state of $H^{\text{SPT}}_m$ is
\begin{equation}   
|\text{GS}\rangle_m=\otimes_{l\in \Lambda} \ket{\mu^x_l=1}\bigotimes U^m(\text{CCZ})\left(\otimes_{j\in \Lambda} \ket{X_j=1}\right),\nonumber
\end{equation}
where
\begin{equation}
\begin{split}
&U(\text{CCZ})=\prod_{(i,j,k)\in\bigtriangleup}\text{CCZ}_{i,j,k}\prod_{(i,j,k)\in\bigtriangledown}\text{CCZ}^{\dagger}_{i,j,k},\nonumber
\end{split}
\end{equation}
with 3-qudit controlled-Z gate acting as
\begin{equation}
    \text{CCZ}_{i,j,k}\ket{s_i,s_j,s_k}=\omega^{s_is_js_k}\ket{s_i,s_j,s_k}.
\end{equation}

\parR{Interface between 2-Rep$(\mathbb Z_4^{(1)}\rtimes \mathbb Z_2^{(0)})$ SPTs.}
{We further present the study of the signature of NISPTs on the interfaces.} 
For example, consider the interface Hamiltonian between  2-Rep$(\mathbb Z_4^{(1)}\rtimes \mathbb Z_2^{(0)})$ SPTs with $m=0$ and $m=1$
\begin{equation}
\begin{split}
    &H_{\text{inter}}=-\sum_l \mu^x_l-\sum_{i\in \text{left}}X_i\\
    &-
   \sum_{i\in \text{right}} X_{i}\prod_{(i,j,k)\in\bigtriangleup} \!\text{CZ}^{\mu}_{i,j,k}\prod_{(i,j,k)\in\bigtriangledown} \!(\text{CZ}^{\mu}_{i,j,k})^{\dagger}+\textrm{(h.c.)},\nonumber
\end{split}
\end{equation}
where $i\in$ left (right) denotes vertices left (right) to the interface $I$ colored by red in Fig.~\ref{fig: interface}. By construction, the ground state of $H_{\text{inter}}$ satisfies $\mu^x_l=1$. Thus the low-energy effective Hamiltonian is 
\begin{equation}
\begin{split}
    &H^{\text{eff}}_{\text{inter}}=-\sum_{i\in \text{left}}X_i\\&-\sum_{i\in\text{right}} X_{i}\prod_{(i,j,k)\in\bigtriangleup} \text{CZ}_{j,k}\prod_{(i,j,k)\in\bigtriangledown} (\text{CZ}_{j,k})^{\dagger}+\textrm{(h.c.)},\nonumber
\end{split}
\end{equation}
which can be simplified by a unitary transformation,
\begin{equation}
\begin{split}
   &U_r(\text{CCZ}) H^{\text{eff}}_{\text{interface}}(U_r(\text{CCZ}))^{\dagger}\\
   =&-\sum_{i\in \text{left}}X_i-\sum_{i\in \text{right}}X_{i}+\textrm{(h.c.)},
\end{split}
\end{equation}
with 
\begin{equation}
U_r(\text{CCZ})=\prod_{(i,j,k)\in\bigtriangleup_r}\text{CCZ}_{i,j,k}\prod_{(i,j,k)\in\bigtriangledown_r}\text{CCZ}^{\dagger}_{i,j,k},\nonumber
\end{equation}
where $\bigtriangleup_r$ and $\bigtriangledown_r$ are the set of triangles on the right side of the interface $I$.

After this transformation, qudits on the interface are all free, resulting in $4^{L_{\text{interface}}}$ degenerate ground states. Moreover, 
 $\mathsf{W}_1$ acts on the ground states localizes around the interface $I$ as
\begin{equation}\label{eq: inter sym} U_r(\text{CCZ})\mathsf{W}_1(U_r(\text{CCZ}))^{\dagger}\ket{\text{GS}}_{\text{inter}}=(U_{\text{inter}}+\textrm{(h.c.)})\ket{\text{GS}}_{\text{inter}},
\end{equation}
where 
\begin{equation}
   U_{\text{inter}} =\prod_{(j,k)\in\bigtriangleup_r\cap I}\!\text{CZ}_{j,k}\prod_{(j,k)\in\bigtriangledown_r\cap I}\!\text{CZ}^{\dagger}_{j,k}\prod_{j\in I} X_j.\nonumber
\end{equation}
In the appendix, we further showed that the above symmetry on the $1d$ interface is anomalous, which forbids the interface from having a unique gapped ground state.

{\parR{Conclusions.}
We introduced the duality method to classify novel NISPTs by mapping gapped phases with $D_{2n}$ symmetry via gauging the nonnormal $\mathbb{Z}_2^{U}$ subgroup. We further constructed the complete family of $\mathrm{Rep}(D_{2n})$-SPTs in $(1+1)D$ and  $2\text{-Rep}(\mathbb{Z}^{(1)}_{4}\rtimes\mathbb{Z}^{(0)}_{2})$-SPTs in $(2+1)D$ on the tensor-product Hilbert space, with a study of the anomalous interfaces. Our results provide a physical understanding of the (higher) fiber functor for (higher) categories. 

Our method can be used for classification and lattice construction of NISPTs with general noninvertible symmetries like $G\times$Rep$(G)$~\cite{Fechisin:2023dkj}, or other gapped phases like noninvertible symmetry-enriched topological orders~\cite{Lu:2025gpt}, which is left to future work. Moreover, because gauging can be realized as quantum operations~\cite{Okada:2024qmk} including measurement~\cite{Tantivasadakarn:2021noi,PhysRevX.14.021040,PhysRevB.109.075116,Lyons:2024fsk}, our method will help prepare NISPTs in quantum devices, e.g., Rydberg atom arrays~\cite{Warman:2024lir}, making noninvertible symmetries routinely testable in both numerical and experimental settings.
}

\parR{Acknowledgments.}We thank Shi Chen, Yunqin Zheng, Yuan Miao, and Yuji Tachikawa for helpful discussions.
W.C.~acknowledges
support from JSPS KAKENHI grant No. JP19H05810, JP22J21553 and JP22KJ1072 and from Villum Fonden Grant no. VIL60714. 
M.Y.~was supported in
part by the JSPS Grant-in-Aid for Scientific Research (Grant No. 20H05860, 23K17689, 23K25865), and
by JST, Japan (PRESTO Grant No. JPMJPR225A, Moonshot R\& D Grant No. JPMJMS2061). 

Towards the completion of this project, we were informed about the forthcoming work~\cite{Bhardwaj:2025piv} on gapped phases with fusion 2-category symmetry in $(2+1)D$. We have coordinated the submissions to arXiv.

\bibliography{ref}

\newpage
\clearpage
\appendix
\widetext
\appendix
\section{Unbroken symmetry of $\mathbb{Z}^U_2$-SSB phase with $D_{2n}$ symmetry}
In this appendix, we find the possible unbroken symmetry group $\Gamma$ of  $\mathbb{Z}^U_2$-SSB phases with $D_{2n}$ symmetry.  These SSB phases should have two-fold degenerate ground states, e.g. spanned by two short-range entangled (SRE) states $|1\rangle$ and $|2\rangle$,  which spontaneously break the $\mathbb{Z}^U_2$ symmetry subgroup.

We first show that the ground states $\ket{1}$ and $\ket{2}$ have the same unbroken symmetry group $\Gamma$ and $\Gamma$ is a normal subgroup of $D_{2n}$ with order $n$. Connected by the $\mathbb Z_2^U$ symmetry operator $U$,  $\ket{1}$ and $\ket{2}$ are orthogonal in the thermodynamic limit. For the ground state $|1\rangle$ with unbroken symmetry group $\Gamma$,  each $g\in \Gamma$ satisfies that $g|1\rangle\propto|1\rangle$. Therefore, 
\begin{equation}
    \langle 2|g|1\rangle=0\quad \to \quad g^{-1}|2\rangle\propto|2\rangle.
\end{equation}
As a result, the unbroken symmetry of $|2\rangle$ is also $\Gamma$. Moreover, for any broken symmetry element $k\notin \Gamma$, 
\begin{equation}
    k|1\rangle\propto|2\rangle,\quad k|2\rangle\propto|1\rangle,
\end{equation}
which result from the fact that $k|1\rangle$ and $k|2\rangle$ are SRE but any linear combination of $|1\rangle$ and $|2\rangle$ is a cat state and thus long-range entangled \footnote{There exits a local order parameter $O$ odd under $\mathbb{Z}^U_2$. It satisfies $\langle 1| O(x) O(y)|1\rangle=\langle 2| O(x) O(y)|2\rangle=\langle 1|O(x)|1\rangle\langle 1|O(y)|1\rangle=\langle 2|O(x)|2\rangle\langle 2|O(y)|2\rangle\ne 0$ when $|x-y|\to \infty$ and $\langle 1|O(x)|1\rangle=-\langle 2| O(x)|2\rangle\ne 0$. Moreover, for any finite region supported operator $O'$, $\langle 2|O'|1\rangle=0$.   Then for any state $|\psi\rangle$ which is a linear combination of $|1\rangle$ and $|2\rangle$, we have $\langle\psi| O(x) O(y)|\psi\rangle\ne \langle\psi| O(x)|\psi\rangle \langle\psi| O(y)|\psi\rangle$ when $|x-y|\to \infty$. Thus $|\psi\rangle$ is long-range entangled. }. From this, we can further show that 
\begin{equation}
    kgk^{-1}|1\rangle\propto|1\rangle,\quad kgk^{-1}|2\rangle\propto|2\rangle,
\end{equation}
i.e., $\Gamma$ is the normal subgroup of $D_{2n}$. 
As there are two ground states, we have $D_{2n}/\Gamma=\mathbb{Z}_2^{U}$. The unbroken symmetry group $\Gamma$ is then an order $n$ normal subgroup of $D_{2n}$ which does not contain the $\mathbb Z_2^U$ generator $U$. In general, there are two cases: 
\begin{enumerate}
    \item The unbroken symmetry group is $\mathbb Z_{n}$ generated by $\eta$.
    \item The unbroken symmetry group involves an element $\eta^m U$. Because $\Gamma$ is the normal subgroup, 
    \begin{equation}
        \eta^{m'} U\eta^m U\eta^{m'} U=\eta^{-m+2m'} U\in G_{\text{unbroken}},\quad \forall m'\in\mathbb{Z}_n.
    \end{equation}
    Hence, we can derive that  
    \begin{equation}
        \eta^m U\eta^{-m+2m'} U=\eta^{-2m'+2m} \in \Gamma,
    \end{equation}
    which implies $\mathbb Z_{n/2}$ generated by $\eta^2$ is unbroken and this can only happen when $n$ is even. Furthermore, if the unbroken symmetry group involves an element $\eta^{2m} U$, then $U$ must be unbroken which conflicts with our set-up. Thus the unbroken symmetry group must be
 $\mathbb Z_{n/2}\rtimes\mathbb Z_2=\{\eta^{2}|\eta^n=1\}\rtimes \{1,\eta U\}$ for $n\in 2\mathbb{Z}$.
\end{enumerate}
In summary, 
\begin{equation}
    \Gamma=
    \begin{cases}
        \mathbb Z_{n},\quad \text{if } n\text{ mod }2=1,\\
        \mathbb Z_n \text{ or } (\mathbb Z_{n/2}\rtimes\mathbb Z_2) ,\quad \text{if } n\text{ mod }2=0.
    \end{cases}
\end{equation}
 
\section{Distinct SPTs by one-site lattice translation}
In this appendix, we will show that when $n=0$ mod 4, the SSB Hamiltonian obtained from \eqref{eq:evenoddphase} by one-site lattice translation is attached with a nontrivial $\mathbb Z_{n/2}^{\eta}\rtimes \mathbb Z_2$-SPT phase. The key feature is that the edge modes carry a projective representation of global symmetry. 

The Hamiltonian under open boundary conditions reads
 \begin{equation}
     H_{\text{OBC}}=-\sum_{k=1}^{2L/n-1}\left(\sum_{j=1}^{n/2-1}\sigma^z_{nk/2-(j-1)}\sigma^z_{nk/2-(j-2)}\right)+\sum_{l=0}^{ n/2-1}\eta^l\left(\sum_{k=1}^{2L/n-2}\sigma^y_{n(k-1)/2+2}\left(\prod_{j=n(k-1)/2+3}^{n(k+1)/2}\sigma^x_{j}\right)\sigma^y_{n(k+1)/2+1}\right)\eta^{-l}.
 \end{equation}
It is easy to find the left-most single spin and the right-most $n/2-1$ spins decouple from this Hamiltonian, which implies the existence of edge modes. For each ground state, the unbroken symmetry thus factorizes into local factors on the boundary as
\begin{equation}
\begin{split}
    &\eta^2|_{\text{GS}}=\eta^2_L \eta^2_R,\quad \eta^2_L=\exp\left(\frac{2\pi i}{n}(1-\sigma^z_1)\right),\quad \eta^2_R=\prod_{j={L-n/2+2}}^L\exp\left(\frac{2\pi i}{n}(1-\sigma^z_j)\right),\,\\  &\eta U|_{\text{GS}}= (\eta U)_L(\eta U)_R,\quad (\eta U)_L=\exp\left(\frac{\pi i}{n}(1-\sigma^z_1)\right)\sigma^x_1, \quad (\eta U)_R=\prod_{j={L-n/2+2}}^L\exp\left(\frac{\pi i}{n}(1-\sigma^z_j)\right)\sigma^x_j.
\end{split}
\end{equation}
On each boundary, these factors form a projective representation of $\mathbb Z_{n/2}^z\rtimes \mathbb Z_2$ since $\eta^{n/2}_{L/R}=(\eta^{2}_{L/R})^{n/4}$ anticommutes with $(\eta U)_{L/R}$.

\section{Projective representation}
In this appendix, we analyze the projective representation at the interfaces between different NISPTs with Rep$(D_{2n})$ symmetry in $(1+1)D$.

First, recall the SSB Hamiltonian \eqref{eq:evenoddphase}. The first term includes all the nearest neighbor interactions within a unit cell of $n/2$ sites and after the KW transformation it becomes
\begin{equation}
    \tilde{H}_1=-\sum_{k=1}^{2L/n}\left(\sum_{j=1}^{n/2-1}\sigma^x_{nk/2-j}\right),
\end{equation}
where the summation includes $\sigma^x$ at every site except the right-most spin in every unit cell.

The second term of the SSB Hamiltonian \eqref{eq:evenoddphase} includes the interactions between nearest unit cells and after the KW transformation it becomes $\tilde{H}_2=\sum_{l=0}^{n/2-1}\tilde{H}_2^{(l)}$ with
\begin{equation}
    \tilde{H}_2^{(0)}\mathsf{W}_{l}=\mathsf{W}_{l}\tilde{H}_2^{(l)},\quad \tilde{H}_2^{(l)}\mathsf{W}_{-l}=\mathsf{W}_{-l}\tilde{H}_2^{(0)}.
\end{equation}
Acting on the ground state, $\tilde{H}_2^{(0)}\ket{\text{GS}}=1$ implies $\tilde{H}_2^{(l)}\ket{\text{GS}}=1$. When we analyze the constraints from the ground state, it is sufficient to focus on the $l=0$ term
\begin{equation}
    \tilde{H}_2^{(0)}=-\sum_{k=1}^{2L/n}\sigma^z_{n(k-1)/2}\left(\prod_{j=n(k-1)/2+1}^{n(k+1)/2-1}\sigma^x_{j}\right)\sigma^z_{n(k+1)/2},
\end{equation}
which includes interaction between three adjacent unit cells. For the $k$-th term, it includes the $\sigma^z$ acting on the right-most spin in the $(k-1)$- and $(k+1)$-th unit cells,  the $\sigma^x$ acting on every spin in the $k$- and $(k+1)$-th unit cells except the right-most spin in the $(k+1)$-th unit cell. It is clear that each term in $\tilde{H}_1$ and $\tilde{H}_2$ commutes with each other. For $n=0$ mod 4, the right-most position in a unit cell in \eqref{eq:evenoddphase} is $nk/2$, an even number. Therefore, this SPT can be referred to as the \textit{even SPT}. And the one after one site translation is the \textit{odd SPT}.

Consider a stabilizer model for the interface Hamiltonian between the even SPT and the product state on a closed chain with $2L$ sites
\begin{equation}
    H_{e|p}=-\sum_{k=1}^{2L/n}\left(\sum_{j=1}^{n/2-1}\sigma^x_{nk/2-j}\right)-\sum_{k=1}^{2L/n-1}\sigma^z_{n(k-1)/2}\left(\prod_{j=n(k-1)/2+1}^{n(k+1)/2-1}\sigma^x_{j}\right)\sigma^z_{n(k+1)/2}+(\text{other part of }\tilde{H}_2)-\sum_{j=L+1}^{2L-1}\sigma^x_{j},
\end{equation}
where the even SPT occupies the first $L$ sites and the $2L$ site, and the product state occupies the remaining sites.
On the ground states, the constraints are
\begin{align}
    \begin{split}
        \sigma^x_{j}&=1,\quad j=L+1,...,2L-1,\\
        \sigma^x_{j}&=1,\quad 1\leq j \leq L \text{ and }j\neq kn/2, \quad  k=1,..,2L/n-1\\
        \sigma^z_{n(k-1)/2}\left(\prod_{j=n(k-1)/2+1}^{n(k+1)/2-1}\sigma^x_{j}\right)\sigma^z_{n(k+1)/2}&=1, \quad  k=1,..,2L/n-1.
    \end{split}
\end{align}
Considering the second constraint, the last one can be simplified as
\begin{equation}
    \sigma^z_{n(k-1)/2}\sigma^x_{nk/2}\sigma^z_{n(k+1)/2}=1, \quad  k=1,..,2L/n-1.
\end{equation}
In total, there are $2L-2$ constraints and the ground states are four-fold degenerate. This indicates the existence of edge modes on the interfaces.

Because of the constraints, the action of $U^{o}$ on the ground states is trivial up to a phase factor. The action of $U^{e}$ on the ground states is
\begin{equation}
    U^{e}\ket{\text{GS}}=\sigma^y_{L}\sigma^y_{2L}\ket{\text{GS}}:=Y^{(\text{L})}Y^{(\text{R})}\ket{\text{GS}}:=U^{e,(\text{L})}U^{e,(\text{R})}\ket{\text{GS}},
\end{equation}
where we label the interface at the $L$-th site as L, and the one at the $2L$-th site as R.
Together with $X^{(\text{R})}:=\sigma^x_{L},X^{(\text{L})}:=\sigma^x_{2L}$, $\{X^{(\text{L}),(\text{R})},Y^{(\text{L}),(\text{R})}\}$ form an operator basis at the interfaces. From the commutation relations
\begin{align}
\begin{split}
    X^{(\text{L}),(\text{R})}\mathsf{W}_l
    &=\mathsf{W}_lX^{(\text{L}),(\text{R})},
\\
   Y^{(\text{L})}Y^{(\text{R})}\mathsf{W}_l
   &=(-1)^{lL}\mathsf{W}_lY^{(\text{L})}Y^{(\text{R})},
\end{split}
\end{align}
and the fusion rule
\begin{equation}
    \mathsf{W}_{n/4}^2=(1+U^o)(1+U^e),
\end{equation}
we have
\begin{align}
    \begin{split}
        X^{(\text{L}),(\text{R})}\mathsf{W}_{n/4}\ket{\text{GS}}&=\mathsf{W}_{n/4}X^{(\text{L}),(\text{R})}\ket{\text{GS}},\\
        Y^{(\text{R})}Y^{(\text{L})}\mathsf{W}_{n/4}\ket{\text{GS}}&=\mathsf{W}_{n/4}Y^{(\text{R})}Y^{(\text{L})}\ket{\text{GS}},\\
        \mathsf{W}_{n/4}^2\ket{\text{GS}}&=2(1+Y^{(\text{R})}Y^{(\text{L})})\ket{\text{GS}}.
    \end{split}
\end{align}
Therefore the effective action of the noninvertible operator $\mathsf{W}_l$ on the ground states is (here we need to fix the overall coefficient by studying the action of $\mathsf{W}$ on the ground states)
\begin{equation}
    \mathsf{W}_{n/4}=X^{\text{(L)}}Z^{\text{(R)}}+Z^{\text{(L)}}X^{\text{(R)}}=\mathsf{W}_{n/4}^{(\text{L}),1}\mathsf{W}_{n/4}^{(\text{R}),1}+\mathsf{W}_{n/4}^{(\text{L}),2}\mathsf{W}_{n/4}^{(\text{R}),2}
\end{equation}
and the projective representation is
\begin{equation}
    \mathsf{W}_{n/4}^{(\text{L}),k}U^{e,(\text{L})}=-U^{e,(\text{L})}\mathsf{W}_{n/4}^{(\text{L}),k},\quad \mathsf{W}_{n/4}^{(\text{R}),k}U^{e,(\text{R})}=-U^{e,(\text{R})}\mathsf{W}_{n/4}^{(\text{R}),k}.
\end{equation}

\section{Gauging in general dimensions}
We first illustrate the gauging method on a general spatial lattice and then focus on the triangular spatial lattice in $(2+1)D$. Consider a general lattice $\Lambda$, in which the vertices are labeled by $i,j$, the edges by $l$, and the faces by $f$. We start from a tensor product Hilbert space on vertices. On each vertex, the local Hilbert space is $\mathbb{C}^2\otimes \mathbb C^n$ , spanned by $\ket{\sigma}_i\otimes\ket{s}_i$ where $\sigma\in \mathbb Z_2$ and $s\in \mathbb Z_n$. Pauli operators $\sigma^x_i,\sigma^z_i$ act on the $\sigma$-spin, while $\mathbb Z_n$ shift and clock operators $X_i,Z_i$ act on the $s$-spin. Further consider the $D_{2n}=\mathbb Z_n^{\eta}\rtimes \mathbb Z_2^U$ symmetry generated by
\begin{equation}\label{eq:D2n generator}
    \eta=\prod_{i\in \Lambda}X_i,\quad U=\prod_{i\in \Lambda} C_i\sigma^x_i,
\end{equation}
where the local charge conjugation operator
\begin{equation}
    C_i=\sum_{s=0}^{n-1}\ket{-s}\bra{s}_i,
\end{equation}
acts only nontrivially on shift and clock operators
\begin{equation}
    C_iX_iC_i^{\dagger}=X_i^{\dagger},\quad  C_iZ_iC_i^{\dagger}=Z_i^{\dagger}.
\end{equation}

Gauging the $\mathbb Z_2^U$ symmetry leads to the dual noninvertible symmetry. 
We introduce $\mathbb Z_2$ gauge field $\mu^x_l,\mu^z_l$ on each edge $l$ and project the extended Hilbert space to gauge invariant sector by imposing Gauss law constraint on every site
\begin{equation}
    G_i=C_i\sigma^x_i\prod_{l\ni i}\mu^z_l=1,\quad \forall i\in \Lambda.
\end{equation}
For convenience, we also impose the flatness condition 
\begin{equation}\label{eq:flatappendix}
    \prod_{l\in f}\mu^x_{l}=1,\quad \forall f\in \Lambda.
\end{equation}
In the dual model, we only consider gauge invariant operators, which commute with the Gauss law operator $G_i$ on every vertex. For example, on any closed loop $\gamma$, there is a $\mathbb Z_2$ operator
\begin{equation}
     W_{\gamma}=\prod_{l\in \gamma}\mu^x_l,
\end{equation}
generating the quantum $(d-1)$-from $\mathbb Z_2$ symmetry in a $d$ dimensional spatial lattice. Because of the flatness condition~\eqref{eq:flatappendix}, $W_{\gamma}$ is nontrivial only when $\gamma$ is a non-contractible loop. We can define the condensation operator
\begin{equation}
    \mathsf{C}=\frac{1}{2^{|\Sigma|}}\prod_{\gamma\in \Sigma}(1+W_{\gamma}),
\end{equation}
where $\Sigma$ is the set of all non-contractible loops in lattice $\Lambda$. On a $2d$ spatial lattice with the periodic boundary condition, the condensation operator is
\begin{equation}
    \mathsf{C}=\frac{1}{4}(1+W_x)(1+W_y).
\end{equation}

On any open path $P_{(i_0,j_0)}$ from vertex $i_0$ to $j_0$, we can define the following gauge invariant operator
\begin{equation}\label{eq:gaugeinvop}
    O_{j_0}^{\prod_{l\in P_{(i_0,j_0)}}\mu^x_{l}}+(\text{h.c.}), 
\end{equation}
where $O$ can be any product of $X,Z$ and 
\begin{equation}
    \mathcal O^{\mu^x_{l}}:=\frac{1+\mu^x_l}{2}\mathcal O+\frac{1-\mu^x_l}{2}\mathcal O^{\dagger}.
\end{equation}
The gauge invariance of \eqref{eq:gaugeinvop} follows from the quasi-gauge-invariant condition
\begin{equation}\label{eq:gaugecovar}
    G_i\left(O_{j_0}^{\prod_{l\in P_{(i_0,j_0)}}\mu^x_{l}}\right)G_i^{\dagger}=
    \begin{cases}
    O_{j_0}^{\prod_{l\in P_{(i_0,j_0)}}\mu^x_{l}},& \quad \forall i\neq i_0,\\
        \left(O_{j_0}^{\prod_{l\in P_{(i_0,j_0)}}\mu^x_{l}}\right)^{\dagger},&\quad i=i_0,
    \end{cases}
\end{equation}
which we show inductively in terms of the length of the path $P_{(i_0,j_0)}$. For the zero-length case, i.e. $i_0=j_0$, by definition we have 
\begin{equation}
    G_iO_{i_0}G_i^{\dagger}=
    \begin{cases}
        O_{i_0},&\quad \forall i\neq i_0\\
        O_{i_0}^{\dagger},&\quad i=i_0.
    \end{cases}
\end{equation}
Suppose \eqref{eq:gaugecovar} is true for any path with length $k$ and let's prove the case of $k+1$. For a path $P_{(i_0,j_0)}$ with length $k+1$, it can be decomposed into a path $P'_{(i'_0,j_0)}$ with length $k$ and a path $P''_{(i_0,i'_0)}$ with length 1 and
\begin{equation}
    O_{j_0}^{\prod_{l\in P_{(i_0,j_0)}}\mu^x_l}=\frac{1+\mu^x_{\braket{i_0,i_0'}}}{2}
    O_{j_0}^{\prod_{l\in P_{(i'_0,j_0)}}\mu^x_l}+\frac{1-\mu^x_{\braket{i_0,i_0'}}}{2}
   \left( O_{j_0}^{\prod_{l\in P_{(i'_0,j_0)}}\mu^x_l}\right)^{\dagger}.
\end{equation}
At site $i_0$, we have
\begin{equation}
    G_{i_0}O_{j_0}^{\prod_{l\in P_{(i_0,j_0)}}\mu^x_l}G_{i_0}^{\dagger}=\frac{1-\mu^x_{\braket{i_0,i_0'}}}{2}
    O_{j_0}^{\prod_{l\in P_{(i'_0,j_0)}}\mu^x_l}+\frac{1+\mu^x_{\braket{i_0,i_0'}}}{2}
    \left(O_{j_0}^{\prod_{l\in P_{(i'_0,j_0)}}\mu^x_l}\right)^{\dagger}=\left(O_{j_0}^{\prod_{l\in P_{(i_0,j_0)}}\mu^x_l}\right)^{\dagger}.
\end{equation}
At site $i_0'$, we have
\begin{equation}
    G_{i_0'}O_{j_0}^{\prod_{l\in P_{(i_0,j_0)}}\mu^x_l}G_{i_0'}^{\dagger}=\frac{1-\mu^x_{\braket{i_0,i_0'}}}{2}
    \left(O_{j_0}^{\prod_{l\in P_{(i'_0,j_0)}}\mu^x_l}\right)^{\dagger}+\frac{1+\mu^x_{\braket{i_0,i_0'}}}{2}
    O_{j_0}^{\prod_{l\in P_{(i'_0,j_0)}}\mu^x_l}=O_{j_0}^{\prod_{l\in P_{(i_0,j_0)}}\mu^x_l}.
\end{equation}
At any other sites, it is straightforward to show that
\begin{equation}
     G_iO_{j_0}^{\prod_{l\in P_{(i_0,j_0)}}\mu^x_l}G_i^{\dagger}=O_{j_0}^{\prod_{l\in P_{(i_0,j_0)}}\mu^x_l}.
\end{equation}
Therefore, we proved the case of length $k+1$ and \eqref{eq:gaugecovar} is true for any length inductively.

The quasi-gauge-invariant condition \eqref{eq:gaugecovar} can be generalized to operator defined on $n$ paths with a common end point $i_0$
\begin{equation}
    O_{j_1}^{\prod_{l\in P_{(i_0,j_1)}}\mu^x_{l}}\cdots O_{j_n}^{\prod_{l\in P_{(i_0,j_n)}}\mu^x_{l}},
\end{equation}
which results in the gauge invariant operator
\begin{equation}
    O_{j_1}^{\prod_{l\in P_{(i_0,j_1)}}\mu^x_{l}}\cdots O_{j_n}^{\prod_{l\in P_{(i_0,j_n)}}\mu^x_{l}}+(\text{h.c.}).
\end{equation}
Following a similar idea, we can obtain the dual operators of the $\mathbb Z_n$ symmetry operators $\eta^k$. When $n$ is odd, there are $(n-1)/2$ zero-form noninvertible operators 
\begin{equation}\label{eq:oddn nisym}
    \mathsf{W}_{k}=\mathsf{C}\left(\prod_{j\in \Lambda} X^{k\prod_{l\in P_{(i_0,j)}}\mu^x_l}_j+(\text{h.c.})\right),
    \quad
    k=1,...,(n-1)/2.
\end{equation}
{Here the condensation operator is given by
\begin{equation}
        \mathsf{C}=\frac{1}{2^{|\Sigma|}}\prod_{\gamma\in \Sigma}(1+W_{\gamma}),
     \end{equation}
where $\Sigma$ is the set of all non-contractible loops.}
This condensation operator ensures the independence of the reference point $i_0$, as we show later in the tensor network representation.

With the convention $\mathsf{W}_0=2\mathsf{C}$ and $\mathsf{W}_{k}=\mathsf{W}_{n-k}$ if $k>(n-1)/2$, the fusion algebra is
\begin{align}\label{eq:evenn nisym}
    \begin{split}
        &\mathsf{W}_{k_1}\times \mathsf{W}_{k_2}=\mathsf{W}_{k_1+k_2}+\mathsf{W}_{k_1-k_2},
        \\
        &W_{\gamma}\mathsf{W}_{k}=\mathsf{W}_{k}W_{\gamma}=\mathsf{W}_{k},\forall \gamma\in \Lambda.
    \end{split}
\end{align}
When $n$ is even, there are $n/2-1$ zero-from noninvertible operators
\begin{equation}
    \mathsf{W}_{k}=\mathsf{C}\left(\prod_{j\in \Lambda} X^{k\prod_{l\in P_{(i_0,j)}}\mu^x_l}_j+(\text{h.c.})\right),
    \quad
    k=1,...,n/2-1
\end{equation}
and one zero-form invertible operator
\begin{equation}
    \eta^{\frac{n}{2}}=\prod_{j\in \Lambda}X_j^{\frac{n}{2}}.
\end{equation}
Further with the convention $\mathsf{W}_{n/2}=2\mathsf C\eta^{\frac{n}{2}}$, the fusion algebra is 
\begin{align}
    \begin{split}
        &\mathsf{W}_{k_1}\times \mathsf{W}_{k_2}=\mathsf{W}_{k_1+k_2}+\mathsf{W}_{k_1-k_2},
        \\
        &W_{\gamma}\mathsf{W}_{k}=\mathsf{W}_{k}W_{\gamma}=\mathsf{W}_{k}.
    \end{split}
\end{align}
If the lattice is $d$-dimensional, we obtain the operators and the fusion algebras for the fusion $d$-category $d$-Rep$(\mathbb Z_n^{d-1}\rtimes \mathbb Z_2^{(0)})$.

\begin{figure}[htp]
	\centering
	\includegraphics[angle=270,origin=c,scale=0.25]{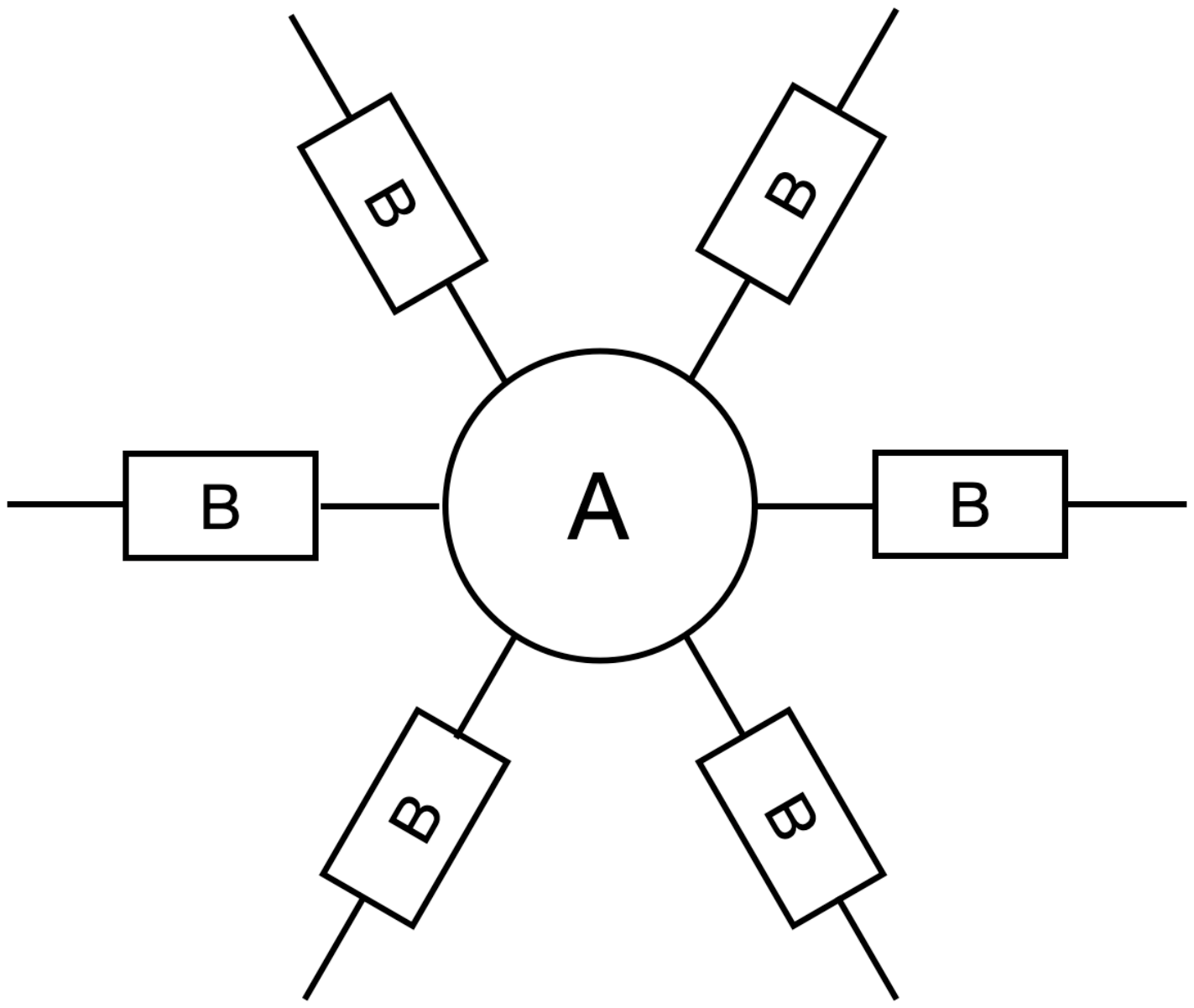}
    \captionof{figure}{ Local tensors on vertices and links. The physical bonds are in the directions perpendicular to the page. $A$-tensor acts on the vertices and $B$-tensor acts on the edges. $A, B$-tensors are connected by virtual bonds.}\label{fig:local tensor}
\end{figure}

Let us turn to the 2-Rep$(\mathbb{Z}^{(4)}_{2}\rtimes \mathbb{Z}^{(0)}_{2})$ symmetry on a 2-dimensional triangular lattice, which can be obtained from zero-form $D_8$ symmetry by gauging the non-normal $\mathbb{Z}^{U}_2$ symmetry. 
The dual model after gauging preserve a quantum $\mathbb{Z}_2^{(1)}$ symmetry, generated by the product of $\mu^x$ on any loop $\gamma$
\begin{equation}
    W_{\gamma}=\prod_{l\in \gamma} \mu^x_l,
\end{equation}
one invertible zero-form symmetry
\begin{equation}
\eta^2=\prod_i X^2_i,
\end{equation}
and one noninvertible zero-form symmetry with $k=1$
\begin{equation}
\begin{split}
    \mathsf{W}_1=\mathsf{C}\left(\prod_{j\in \Lambda} X^{\prod_{l\in P_{(i_0,j)}}\mu^x_l}_j+(\text{h.c.})\right), \quad \mathsf{C}=\frac{1}{4}(1+W_x)(1+W_y).
    \end{split}
\end{equation}
Symmetry operators $\eta^2$, $W_{\gamma}$ and $\mathsf{W}_1$ generate the fusion algebra of the fusion 2-category 2-Rep$(\mathbb Z_4^{(1)}\rtimes \mathbb Z_2^{(0)})$ 
 \begin{equation}
\begin{split}
    &\mathsf{W}_1^2=2\mathsf{C}(1+\eta^2), \quad \eta^2W_{\gamma}=W_{\gamma}\eta^2\\
    &\eta^2\mathsf{W}_1=\mathsf{W}_1\eta^2=W_{\gamma} \mathsf{W}_1=\mathsf{W}_1 W_{\gamma}=\mathsf{W}_1,\quad \forall \gamma\in \Lambda.
    \end{split}
\end{equation}
The noninvertible operator $\mathsf{W}_1$ does not depend on the reference point $i_0$ and has a local representation. At first, we notice that the condensation operator $\mathsf C$ and flat gauge field condition projects to the subspace satisfying $W_x=W_y=1$ and $\prod_{l\in f}\mu^x_f=1$ for each triangle $f$. This subspace can be spanned by the states $|\{\mu^x_{\langle j_1,j_2\rangle}=\tau_{j_1} \tau_{j_2}\}\rangle$ where $\langle j_1,j_2\rangle$ is a link connecting vertices $j_1,j_2$ and $\tau_j=\pm 1$ is defined on each vertex.  Hence, we can rewrite $\mathsf W_1$ as
\begin{equation}
\begin{split}
    \mathsf{W}_1&=\frac{1}{2}\sum_{\{\tau_j=\pm 1\}}\left(\prod_{j\in \Lambda} X^{\tau_{i_0}\tau_j}_j+(\text{h.c.})\right)\prod_{\langle j_1,j_2\rangle\in \Lambda} \delta^{\mu^x_{\langle j_1,j_2\rangle}}_{\tau_{j_1} \tau_{j_2} }
    \\&=\sum_{\{\tau_j=\pm 1\}}\prod_{j\in \Lambda} X^{\tau_j}_j\prod_{\langle j_1,j_2\rangle\in \Lambda} \delta^{\mu^x_{\langle j_1,j_2\rangle}}_{\tau_{j_1} \tau_{j_2} }, 
    \end{split}
\end{equation}
where $\delta^{\mu^x_{\langle j_1,j_2\rangle}}_{\tau_{j_1} \tau_{j_2}}$ is a projection operator ensuring $\mu^x_{\langle j,j'\rangle}=\tau_j \tau_j'$ and $1/2$ in the first equation comes from the one-to-two correspondence between $\mu^x$ and $\tau$ configurations. This local representation does not depend on the starting point of the path $P_{(i_0,j)}$. Moreover, it can be presented in the tensor network language. The $\tau$ degree of freedom can be understood as the virtual degree of freedom in local tensor with bond dimension 2, which is shown in Fig.~\ref{fig:local tensor}.
On vertex $j$ connected with vertual bonds $n_1,n_2,n_3,n_4,n_5,n_6$, $A$-tensor is
{
\begin{equation}
    A_j^{n_1n_2n_3n_4n_5n_6}=X_j^{\tau_j}\prod_{i=1}^{6}\delta_{\tau_j}^{n_i}.
\end{equation}
}
On edge $l$ connecting the virtual bonds $n_1,n_2$, the $B$-tensor is
\begin{equation}
    B^{n_1n_2}_l=\ket{\mu^x_l}\bra{\mu^x_l}, \quad {\mu^x_l=n_1 n_2}.
\end{equation}

This representation can be generalized to 0-form noninvertible operators \eqref{eq:evenn nisym} and \eqref{eq:oddn nisym}  on a general spatial lattice $\Lambda$
\begin{equation}
\begin{split}
    \mathsf{W}_k&=\sum_{\{\tau_j=\pm\}}\prod_{j\in \Lambda} X^{k\tau_j}_j\prod_{\langle j_1,j_2\rangle\in \Lambda} \delta^{\mu^x_{\langle j_1,j_2\rangle}}_{\tau_{j_1} \tau_{j_2} }.
    \end{split}
\end{equation}

\section{Anomalies of the interface between stacked SPT phases before gauging}
In this appendix, we will determine SPT phases stacked on SSB phases before gauging, by studying the symmetry action on the interface between SPT phases and trivial phases, and calculating the anomaly index using Else-Nayak approach~\cite{Else:2014vma,Wang:2024vjl,PhysRevB.106.224420}. 

\subsection{Review of Else-Nayak approach}
We first review the Else-Nayak approach in $(1+1)D$. For a symmetry group $G$ with group element $g$, we can define unitary symmetry operator $U(g)$ acting on the spacial direction, and its restriction to a subregion $M$, $U_M(g)$. $U_M$ and $U$ are the same in the interior of $M$, but are ambiguous up to some local unitaries near the boundary $\partial M$. We assume each symmetry operator is finite-depth local unitary and then we have 
\begin{equation}
U_M(g_1)U_M(g_2)=W(g_1,g_2)U_M(g_1g_2),
\label{eq:W}
\end{equation}
for some local unitary $W(g_1,g_2)$ acting on $\partial M$. The associativity of $U_M(g_1)U_M(g_2)U_M(g_3)$ dictates that $W$ should satisfy
\[\label{eq:associativity pure state}
W(g_1,g_2)W(g_1g_2,g_3)=^{U_M(g_1)}W(g_2,g_3)W(g_1,g_2g_3),
\]
where we use the shorthand notation $^{U_M}W\equiv U_M W U_M^{-1}$. 
Then we further restrict $W$ to the left and right end of $M$:
\begin{equation}
W(g_1,g_2)=W_l(g_1,g_2)W_r(g_1,g_2).
\label{eq:Wl}
\end{equation}
$W_l$ and $W_r$ satisfies the consistent condition \eqref{eq:associativity pure state} up to a phase factor $\omega$:
\[
\begin{split}
&W_{l}(g_1,g_2)W_l(g_1g_2,g_3)\\
&=\omega(g_1,g_2,g_3)^{U_M(g_1)}W_l(g_2,g_3)W_l(g_1,g_2g_3),\\
&W_{r}(g_1,g_2)W_r(g_1g_2,g_3)\\
&=\omega^{-1}(g_1,g_2,g_3)^{U_M(g_1)}W_r(g_2,g_3)W_r(g_1,g_2g_3).
\end{split}
\label{eq:omega}
\]
From the associativity relation of $W_l(g_1,g_3)W_l(g_1g_2,g_3)W_l(g_1g_2g_3,g_4)$, one can show that $\omega$ must satisfy the following 3-cocycle condition and 
$\omega$ is only uniquely defined modulo a 2-coboundary, i.e.,
\begin{equation}
\begin{split}
\omega(g_1,g_2,g_3)\sim \omega(g_1,g_2,g_3)\beta(g_1,g_2)\beta(g_1g_2,g_3)\beta^{-1}(g_2,g_3)\beta^{-1}(g_1,g_2g_3).
\label{eq:coboundary pure state}
\end{split}
\end{equation}
Consequently, the anomaly of $G
$ in $(1+1)D$ can be classified by $[\omega^G]$, where $[\cdot]$ denotes an equivalence class defined in \eqref{eq:coboundary pure state}, representing an element in the third cohomology group $H^3(G, U(1))$.
\subsection{Interface anomalies of $\mathbb{Z}^{\eta}_4$-SPT with $m=1$}
We start with the Hamiltonian \eqref{eq:2Dungauged Hal} and show the $\mathbb Z_4$ qudit is in the SPT phases of $\mathbb Z^{\eta}_4$ distinguished by $m$.   We consider the interface Hamiltonian of qudits between $m=0$ and $m=1$:
\begin{equation}
\begin{split}
    H_{0-1}=-\sum_{i\in \text{left}}X_i-\sum_{i\in \text{right}} X_{i}\left(\prod_{(i,j,k)\! \in\bigtriangleup} \text{CZ}_{j,k}\prod_{(i,j,k)\!\in\bigtriangledown} (\text{CZ}_{j,k})^{\dagger}\right)^m+ \textrm{(h.c.)}.
\end{split}
\end{equation}
with the interface $I$ colored by red in Fig.~\ref{fig: interface}.

To show  the interface DOF, we consider the unitary $U(\text{CCZ})$ transformation:
\begin{equation}
\begin{split}
    U(\text{CCZ})  H_{0-1} U^{\dagger}(\text{CCZ})=-\sum_{\langle i,j\rangle\in {\Lambda
    }}\sigma^z_i\sigma^z_j-\sum_{i\in \text{left}}X_i-\sum_{i\in \text{right}} X_{i}+ \textrm{(h.c.)}.
\end{split}
\end{equation}
After transformation, the qudits on the interface are all free. We can further determine the
symmetry action on these qudits  \begin{equation}
\begin{split}
 U(\eta)= U(\text{CCZ})  \eta U^{\dagger}(\text{CCZ})|_{\text{GS}}=\left(\prod_{(j,k)\in\bigtriangleup_r\cap I}\text{CZ}_{j,k}\prod_{(j,k)\in\bigtriangledown_r\cap I}\text{CZ}^{\dagger}_{j,k}\right)\prod_{j\in I} X_j.
\end{split}
\end{equation}
 This symmetry action is precisely $U_{\text{inter}}$.

To show the existence of an anomaly of the symmetry on the interface, we first rewrite it as
\begin{equation}
  U(\eta)=  \prod^{L_{\text{inter}/2}}_j \text{CZ}_{2j-1,2j}\text{CZ}^{\dagger}_{2j,2j+1}\prod^{L_{\text{inter}}}_j X_j.
\end{equation}
Then we apply the Else-Nayak approach
to determine its anomaly cocycle. 
Let us restrict this $\mathbb{Z}_4$ symmetry operators to a subregion M
\begin{equation}
\begin{split}
   & U_{M}(\eta)=\prod^{k}_{j=i} \text{CZ}_{2j-1,2j}\text{CZ}^{\dagger}_{2j,2j+1}\prod^{2k+1}_{j=2i-1} X_j,
     \\
     &U_{M}(\eta^2)=(\prod^{k}_{j=i} \text{CZ}_{2j-1,2j}\text{CZ}^{\dagger}_{2j,2j+1})^2\prod^{2k+1}_{j=2i-1} X^{2}_j.
 \end{split}
\end{equation}
 It is easy to check 
 \begin{equation}
 \begin{split}
&W_l(\eta^a,\eta^b)=Z^{ab}_{2i-1},\quad a,b=0,1,2,3.
      \end{split}
 \end{equation}
Hence we can obtain
\begin{equation}
    \omega(\eta^a,\eta^b,\eta^c)=i^{abc},\quad a,b,c=0,1,2,3.
\end{equation}
Then we can calculate the coboundary-invariant combination:
\begin{equation}
\omega(\eta,\eta,\eta)\omega^{-1}(\eta^3,\eta^3,\eta^3)\omega^{-1}(\eta^2,\eta,\eta^2)=-1.
\end{equation}
Since the ground state with $m=0$ is the product state which is in the trivial phase, this minus sign implies the corresponding SPT phase with $m=1$ is nontrivial. More precisely, this SPT phase corresponds to the generator of $\mathbb Z_2$ subgroup of $\mathbb Z_4$ classification \cite{PhysRevB.91.035134,PhysRevLett.114.031601}. 

\section{Proof of anomalous symmetry on the interface}\label{app:anomalous interface}
To study anomaly of symmetry on the interface \eqref{eq: inter sym}, we first rewrite $U_{\text{inter}}$ as
\begin{equation}
    U_{\text{inter}}=\prod^{L_{\text{inter}}}_j X_j \prod^{L_{\text{inter}/2}}_j \text{CZ}_{2j-1,2j}\text{CZ}^{\dagger}_{2j,2j+1},
\end{equation}
where the links $\langle2j-1, 2j\rangle$ and $\langle2j, 2j+1\rangle$ are on the up and down triangles on the right side, respectively. 
Next, because this symmetry operator does not involve translation, we assume the $1d$ symmetric Hamiltonian on the interface can be written as 
\begin{equation}
    H_{1d}=-\sum_{k} H_k,
\end{equation}
 where $H_k$ is a local term that is supported on a finite region $M_k$  and commutes with $(U_{\text{inter}}+U^{\dagger}_{\text{inter}})$. Then we have
 \begin{align}
     \begin{split}
         &H_k(U_{\text{inter}}+U^{\dagger}_{\text{inter}})=(U_{\text{inter}}+U^{\dagger}_{\text{inter}})H_k\\
         \to\quad  &H_kU_{\text{inter}}-U_{\text{inter}}H_k=U^{\dagger}_{\text{inter}}H_k-H_kU^{\dagger}_{\text{inter}}\\
         \to \quad & H_k-U_{\text{inter}}H_kU^{\dagger}_{\text{inter}}=(U^{\dagger}_{\text{inter}}H_k-H_kU^{\dagger}_{\text{inter}})U^{\dagger}_{\text{inter}}\\
         \to \quad & H_k- U_{\text{inter}}H_{k}U^{\dagger}_{\text{inter}}=(U^{\dagger}_{\text{inter}}H_kU_{\text{inter}}-H_k)U^2_{\text{inter}}.
     \end{split}
 \end{align}
Since $U_{\text{inter}}$ is a finite-depth local circuit,  the left side term is indeed local and it is supported on a finite region $N_k$ which we can assume spans from the site $2m$ to $2n+1$.  In contrast, the right side term is highly non-local, potentially acting on qudits far beyond the region $N_k$.  More precisely, we can expand the equation above in the $Z$-basis wave function. We choose  $|\psi_a\rangle=|\{s^a_{j\in N_k}\}\rangle\otimes|\{s^a_{j\in \bar{N}_k}\}$  and $|\psi_b\rangle=|\{s^b_{j\in N_k}\}\otimes|\{s^b_{j\in \bar{N}_k}\}\rangle$ where $\bar{N}_k$ is the
complement of $N_k$ in the $1d$ interface and $s^a_{j\in \bar{N}_k}=s^b_{j\in \bar{N}_k}$. This yields
  \begin{equation}
  \begin{split}
       \text{Left}&= \langle\psi_a|(H_k- U_{\text{inter}}H_{k}U^{\dagger}_{\text{inter}}) |\psi_b\rangle  =  \langle\{s^a_{j\in N_k}\}|(H_k- U_{\text{inter}}H_{k}U^{\dagger}_{\text{inter}}) |\{s^b_{j\in N_k}\}\rangle 
       \\
       \text{Right}&=\langle\psi_a|(U^{\dagger}_{\text{inter}}HU_{\text{inter}}-H)U^2_{\text{inter}} |\psi_b\rangle\\
       &=\langle\{s^a_{j\in N_k}\}|(U^{\dagger}_{\text{inter}}H_{k}U_{\text{inter}}-H_k) |\{s^b_{j\in N_k}\}\rangle (-1)^{\sum^{j=2n+1}_{j=2m+1}s^b_{j-1}s^b_j}(-1)^{\sum_{j<2m+1 }s^b_{j-1}s^b_j+\sum_{j>2n+1 }s^b_{j-1}s^b_j}.
       \end{split}
 \end{equation}
  
 The left side term only depends on qudits belonging to $N_k$, but the right side term also depends on qudits belonging to $\bar{N}_k$. For example,  we can consider two configurations $\{s^{b_1}_j\}$ and $\{s^{b_2}_j\}$ where four adjacent qudits in $\bar{N}_k$ satisfying $s^{b_1}_{j}=s^{b_1}_{j+1}=s^{b_1}_{j+2}=s^{b_1}_{j+3}=0$ or $s^{b_2}_{j}=s^{b_2}_{j+3}=0, s^b_{j+1}=s^b_{j+2}=1\quad (j>2n+1)$ and the other qudits $s^{b_1}_{j}=s^{b_2}_j$.  Then the resulting right side terms from these two configurations will differ by $-1$ sign while the resulting left-side terms are the same. The only consistent solution is for both sides to vanish, implying that $H_k$ commute with $ U_{\text{inter}}$. Hence, any  local Hamiltonian commuting with $(U_{\text{inter}}+U^{\dagger}_{\text{inter}})$ will also commute with $U_{\text{inter}}$. As we prove in the previous appendix that $ U_{\text{inter}}$ is an anomalous $\mathbb{Z}_4$ symmetry, it follows that such a symmetric Hamiltonian can not have a unique ground state.

 \section{The classification based on the conjugacy classes of $\Gamma$-SPTs}
 In this appendix, we provide the calculation on classification of the noninvertible SPT phases, which is equivalent to that of $\Z^U_2$-SSB phases in the dual $D_{2n}$ symmetric systems. In particular, we will focus on the conjugacy classes of the SPT phases of unbroken subgroup $\Gamma$, which is induced by the broken symmetry $\Z^U_2$, as the $\Gamma$-SPT preserves $\Gamma$ symmetry. A similar calculation for $S_3\times \Z_3$ symmetry in $d=1$ is presented in \cite{Aksoy:2025rmg}.
 
 Let us first denote the two ground states of a $\Z^U_2$-SSB phase as  $(\Omega^{\Gamma}_{k}, \Omega^{\Gamma}_{k'})$ where $\Omega^{\Gamma}_{k}$ and $\Omega^{\Gamma}_{k'}$ labels the SPT phase of $\Gamma$ symmetry with level $k$ and $k'$. Due to the broken symmetry $\Z^U_2$, the two ground states should be connected by $U$ and we have $\Omega^{\Gamma}_{k'}=U\Omega^{\Gamma}_{k}$. According to \cite{Aksoy:2025rmg}, the classification of $\Z^U_2$-SSB phase is the conjugacy classes of $\Omega^{\Gamma}_{k}$.
 Physically, this is because the SSB phase with ground states $(\Omega^{\Gamma}_{k}, U\Omega^{\Gamma}_{k})$ is in the same phase as the SSB phase with ground states $(U\Omega^{\Gamma}_{k}, \Omega^{\Gamma}_{k})$:
 \[\label{eq:equivalence relation}
 (\Omega^{\Gamma}_{k}, U\Omega^{\Gamma}_{k})\sim (U\Omega^{\Gamma}_{k}, \Omega^{\Gamma}_{k}).
 \]
 Thus, in each unbroken subgroup branch, the $\Z^U_2$-SSB are classified by the $\Gamma
 $-group cohomology modulo the above equivalence relation induce by $U$, which is equivalent to the conjugacy classes in Eq.~\eqref{eq:conjugacy classes}. 
 For example, when $n=4$ in $(2+1)D$ with lattice realization in Eq.~\eqref{eq:lattice D8 sym} and Eq.~\eqref{eq:2Dungauged Hal}, the two ground states of $\Z^U_2$-SSB phase can be realized with all $\sigma$-spin up and down and the  $\Gamma
 $-SPT is characterized by qudits. These two ground states can be flipped by $\prod_j \sigma^x_j$, a $D_8$ symmetric finite-depth circuit. This implies that the systems before and after flipping are in the same $D_8$ phase. This statement can be easily generalized to $\Z_n$-qudits (together with a qubit) on general lattices , where $D_{2n}$ operators are realized as eq.\eqref{eq:D2n generator}.

  The equivalence relation \eqref{eq:equivalence relation} is nontrivial when the broken symmetry $\Z^U_2$ acts nontrivially on the $\Gamma$-SPT phases. This nontrivial action of $U$ comes from the fact that $U$ is not the center of $D_{2n}$ since it acts on the generator of $\Gamma$ in the following ways
\begin{enumerate}
    \item When $\Gamma=\Z^{\eta}_n$, we have
    \[
    U: \eta \to \eta^{-1}.
    \]
    
    \item When $\Gamma=\Z^{\eta^2}_{n/2}\rtimes \Z^{\eta U}_2$, the generators $\eta^2,\eta U$ are mapping to
    \[\label{eq:H trans2}
    U: \eta^2\to \eta^{-2}, \eta U\to \eta^{-1} U=\eta^{-2}\eta U.
    \]

\end{enumerate}
In the following section, we will compute the classification of $\Z^U_2$-SSB phases, which is equivalent
to the classification of dual NISPT phases,  for $d=1,2,3$.
 \subsection{One spatial dimension}
 Let us warm up with one spatial dimensional cases. 
 
 When $n$ is odd, the unbroken subgroup $\Gamma$ can only be $\Z_n$. There is only one trivial SPT phase and it is invariant under $U$. 
 
 When $n$ is even, $\Gamma$ can be $\Z^{\eta}_n$ or $\mathbb Z_{n/2}^{\eta^2}\rtimes\mathbb Z_2^{\eta U}=\{\eta^2|\eta^n=1\}\rtimes \{1,\eta U\}$. In the first case where $\Gamma=\Z^{\eta}_n$, there is also one trivial SPT phase which is invariant under $U$. In the second case where $\Gamma=\mathbb Z_{n/2}^{\eta^2}\rtimes\mathbb Z_2^{\eta U}$, we have $H^2(\mathbb Z_{n/2}^{\eta^2}\rtimes\mathbb Z_2^{\eta U}, U(1))=\mathbb Z_{\text{gcd}(n/2,2)}$. When  $n=2$ mod 4, there is also one trivial SPT phase which is invariant under $U$. When $n=0$ mod 4, two different SPTs can be classified by the projective representation $U_{e}(h)$ with $h\in \Gamma$ on the edge modes, which is determined by the choice of $\pm 1$ in the (anti)-commutation relation:
 \[\label{eq:commu relation1}
 U_{e}(\eta^{2m}) U_e(\eta U)=(\pm 1)^m  U_{e}(\eta U)U_e(\eta^{-2m}), \quad U_e(\eta^{2m'}) U_e(\eta^{2m''}U^k)= U_e(\eta^{2m'+2m''}U^k), 
 \]
 for $m,m', m''\in \Z_{n/2}$ and $k\in\Z_2$.
 For example, the case $U_e(\eta^2)=\sigma^z$ and $U_e(\eta U)=\sigma^x$ can produce the negative phase. Then if we act $U$ on the SPTs, the new projective representation is $U'_e(\eta^{2m})=U_e(\eta^{-2m})$ and $U'_e(\eta U)=U_e(\eta^{-1} U)$ . Thus (anti)-commutation relation is changed to 
 \[\label{eq:commu relation2}
 U'_e(\eta^{-2m}) U'_e(\eta^{-1} U)=(\pm 1)^m  U'_e(\eta^{-1} U)U'_e(\eta^{2m}) , \quad U'_e(\eta^{-2m'}) U'_e(\eta^{-2m''}U^k)= U'_e(\eta^{-2m'-2m''}U^k).
 \]
It is straightforward to find that \eqref{eq:commu relation1} is the same as \eqref{eq:commu relation2}  if we do the mapping: $m\to -m-1$ and $m',m''\to -m',-m''$. Hence the (non-)trivial SPT with $+$ ($-$) sign will remain invariant under $U$.

In summary, the conjugacy classes in Eq.~\eqref{eq:conjugacy classes} are trivial in $(1+1)D$ and all $\Gamma$-SPTs are invariant under $U$.

\subsection{Two spatial dimensions}
In this section, we extend the discussion above to two spatial dimensions.

When $n$ is odd, the unbroken symmetry $\Gamma=\Z^{\eta}_n$ and the $\Gamma$-SPT phase is classified by $H^3(\Z^{\eta}_n, U(1))=\Z_n$. The SPT with the level $k\in\Z_n$ can be classified by the topological response  action \cite{PhysRevB.91.035134,PhysRevLett.114.031601}:
 \[\label{eq: Zncocycle}
 \exp \left(\frac{2\pi  ik}{n^2}\int_{M_3}A dA\right)
 \]
 where $A$ is $\Z_n$ background gauge field satisfying $\oint A\in\Z_n$ for any non-contractible loop and $\oint dA=0$ (mod $n$) for any  non-contractible closed surface. And $M_3$ is a three dimensional manifold. 
Under the $U$ transformation, $A$ is mapped to $-A$ and thus the $\Gamma$-SPT phase is invariant under $U$. Hence there are $n$ different $\Z^U_2$-SSB phases when $n=1$ (mod $2$). This result also implies the $\Z_n$-SPTs always has an larger $D_{2n}$ symmetry, which belongs to $D_{2n}$-SPTs, and we will use this statement in the following discussion. 

When $n$ is even, $\Gamma$ can be $\Z_n$ or $\mathbb Z_{n/2}^{\eta^2}\rtimes\mathbb Z_2^{\eta U}=\{\eta^2|\eta^n=1\}\rtimes \{1,\eta U\}$. In the first case where $\Gamma=\Z^{\eta}_n$, due to the same reason above, the $\Z_n$-SPT phase is invariant under $U$ transformation.  In the second case where $\Gamma=\mathbb Z_{n/2}^{\eta^2}\rtimes\mathbb Z_2^{\eta U}$, the $\Gamma$-SPT phase is classified by
\begin{equation}\label{eq: Dnclass}
H^{3}(\mathbb{Z}^{\eta^2}_{n/2}\rtimes \mathbb{Z}^{\eta U}_2, U(1)) =
\begin{cases}
\Z_{n/2}\times \Z_2\times \Z_2 &\quad\text{$n/2$ is even},\\
\Z_{n/2}\times \Z_2&\quad\text{$n/2$ is odd}. \\
\end{cases}
\end{equation}
Unfortunately, there is no result for the corresponding response theory to our knowledge.  However, in the following section, we will use some simple facts to determine the conjugate classes.

We first consider $n/2$ is odd. 
We can label the SPT phase using the index $(k_1, k_2)$ where $k_1\in\Z_{n/2}$ and $k_2\in \Z_2$. Then we denote the index of SPT after $U$ transformation is $(f_1(k_1,k_2),f_2(k_1,k_2))$. Since the group-SPT phase has the stacking structure, $f$ should a linear function, i.e. $(f_1(k_1+k'_1,k_2+k'_2),f_2(k_1+k'_1,k_2+k'_2))=(f_1(k_1,k_2)+f_1(k'_1,k'_2),f_2(k_1,k_2)+f_2(k'_1,k'_2))$. Thus $f_i(k_1,k_2)=f_{i1} k_1+f_{i2}k_2$.
Here are some facts which can help us identify the result. 
\begin{enumerate}
    \item Firstly, when $k_2=0$, the SPT phases with $\Z_{n/2}$ classification are indeed distinct  $\Z^{\eta^2}_{n/2}$-SPTs with the similar topological response action \eqref{eq: Zncocycle}. This is because this type of topological action is invariant under $\eta U$, which maps $A\to -A$, and thus enjoys the full $\mathbb Z_{n/2}^{\eta^2}\rtimes\mathbb Z_2^{\eta U}$ symmetry. Moreover, as mentioned before, such SPT phase is also invariant under the $U$ transformation. Hence we have $f_{11}=1$. 
    \item  Moreover, since $k_1$ has odd periodicity and $k_2$ has even periodicity, we have $f_{12}=f_{21}=0$. 
\item At last, since $U$ is invertible transformation, we have $f_2(0,1)=1$ and thus $f_{22}=1$. 
    \end{enumerate}
    Thus we prove  the $\mathbb Z_{n/2}^{\eta^2}\rtimes\mathbb Z_2^{\eta U}$-SPT phase is invariant under $U$ when $n/2$ is odd. Combing with case where $\Gamma=\Z_n^{\eta}$, there are $n+n=2n$ different $\Z^U_2$-SSB phases when $n=2$ (mod $4$).

Now let us further consider the case  $n/2$ is even. There is a extra $\Z_2$ in eq.\eqref{eq: Dnclass} which labels the mixed SPT. We use the index $(k_1,k_2,k_3)$ and $(f_1,f_2,f_3)$ to label the level before and after $U$ transformation. We consider the following simple facts. 
\begin{enumerate}
    \item Firstly, when $k_2=k_3=0$, by the same reason above, we can determine $f_{11}=1$.
    \item Next, we consider the SPT solely protected by $\Z^{\eta U}_2$ symmetry where $k_2=1,k_1=k_3=0$. This implies $\Z^{\eta^2}_{n/2}$ is anomaly-free but $\Z^{\eta U}_2$ is anomalous on the boundary. Such anomalous symmetry can be realized by the symmetry operators 
    \[
    U_e (\eta^2)=\prod^L_{j=1} X_j, \quad U_e (\eta U)=\prod^L_{j=1} C_j \prod^L_{j=1} \sigma^x_j \prod^L_{j=1}\text{CZ}^{\sigma}_{j,j+1}.
    \]
 Here on each vertex, we assign a qubit and a $\mathbb Z_{n/2}$ qudit. $X$ is the shift operator of $\Z_{n/2}$ qudits, $C$ is charge conjugation operator, and $\text{CZ}^{\sigma}_{j,j+1}$ is the controlled-Z gate of two qubits. Indeed the anomaly of $\Z^{\eta U}_2$ fully comes from the non-onsite realization on the qubits \cite{PhysRevB.84.235141}.  Then after $U$ transformation, the mapping of generators is
 \[
 U'_e (\eta^2)=U_e (\eta^{-2})=\prod^L_{j=1} X^{-1}_j,\quad U'_e (\eta U)=U_e (\eta^{-1}U)= U_e(\eta^{-2}) U_e (\eta U)=\prod_j X^{-1}_j \prod^L_{j=1} C_j \prod^L_{j=1} \sigma^x_j \prod^L_{j=1}\text{CZ}^{\sigma}_{j,j+1}. 
 \]
 It is straightforward to show the $U'_e (\eta^2)$ is anomaly free but $U'_e (\eta U)$ is anomalous due to the same non-onsite realization on the qubits. This result implies $f_1(0,1,0)=0$ and $f_2(0,1,0)=1$, which help us determine $f_{12}=0, f_{22}=1$. 
    \item Moreover, we consider the mixed SPT with $k_1=k_2=0$ and $k_3=1$, whose boundary anomalous symmetries can be constructed as follows. 
    \[
    U_e(\eta^2)=\prod^L_{j=1} X_j\sigma^x_j, \quad U_e(\eta U)=\prod^L_{j=1} C_j \prod^L_{j=1}\text{CZ}^{\sigma}_{j,j+1}.
    \]
     One can show that both $\Z^{\eta^2}_{n/2}$ and $\Z^{\eta U}_{2}$ are anomaly free as they allows symmetric product states. But due to the discussion above, $\Z_2$ symmetry generated by $U_e(\eta^{-2})U_e(\eta U)$ is anomalous. This implies there is a mixed anomaly between $\Z^{\eta^2}_{n/2}$ and $\Z^{\eta U}_{2}$. Then if we apply $U$, we have $U'_e (\eta^2)=U_e (\eta^{-2})$ and $U'_e (\eta U)=U_e (\eta^{-1}U)= U_e(\eta^{-2}) U_e (\eta U)$. It is straightforward to obtain that the $U'_e (\eta^2)$ is anomaly free but $U'_e (\eta U)$ is anomalous. Hence we have $f_{13}=0, f_{23}=1$.This implies $f_1=k_1, f_2=k_2+k_3+f_{21}k_1 (\text{mod}~ 2), f_3=f_{33}k_3+f_{31}k_1+f_{32}k_2 (\text{mod}~ 2)$ and all remaining $f_{ij}=0,1$.
    \item Finally, we note the $U$ transformation is a $\Z_2$ map. Hence we have $f_{i}(f_1,f_2,f_3)=k_i$. In particular
\begin{equation} f_2(f_1,f_2,f_3)=f_2+f_3+f_{21}f_1=k_2+k_3+f_{21}k_1+f_{33}k_3+f_{31}k_1+f_{32}k_2+f_{21}k_1=k_2 ~(\text{mod}~ 2)
\end{equation}
This constraint tell us that $f_{31}=0, f_{32}=0, f_{33}=1$.
Hence we have $f_1=k_1, f_2=k_2+k_3+f_{21}k_1 ~(\text{mod}~ 2), f_3=k_3$.
\end{enumerate}
 Indeed no matter $f_{21}$ is zero or one, there are $n$ $\Gamma$-SPTs invariant under the $U$ transformation in both cases due to
 \begin{enumerate}
     \item If $f_{21}=0$, the invariant SPT satisfies $k_3=0$.
     \item If $f_{21}=1$, the invariant SPT satisfies $k_1=k_3 ~(\text{mod}~ 2)$.
 \end{enumerate}
On the other hand, $U$ acts nontrivially on the remaining $n$ $\Gamma$-SPTs and there will be $n/2$ pairs. 

In summary, combing with case where $\Gamma=\Z_n^{\eta}$, there are $n/2+n+n=5n/2$ different $\Z^U_2$-SSB phases when $n=0$ (mod $4$). 

\subsection{Three spatial dimensions}
Finally, we will discuss the classification in three spatial dimension.

When $n$ is odd, $\Gamma=\Z^{\eta}_n$ and there is only one trivial SPT phase which is invariant under $U$.

When $n$ is even, $\Gamma$ can be $\Z_n$ or $\mathbb Z_{n/2}^{\eta^2}\rtimes\mathbb Z_2^{\eta U}=\{\eta^2|\eta^n=1\}\rtimes \{1,\eta U\}$. In the first case, there is also one trivial SPT phase which is invariant under $U$. In the second case, we have 
\[
H^4(\mathbb Z_{n/2}^{\eta^2}\rtimes\mathbb Z_2^{\eta U}, U(1))=\mathbb Z^2_{\text{gcd}(n/2,2)}.
\]
When  $n=2$ mod 4, there is also one trivial SPT phase which is invariant under $U$. Thus, combing with the case $\Gamma=\mathbb{Z}_n$, there are $2$ distinct $\Z^U_2$-SSB phase.

When $n=0$ mod 4, there are four different $\Gamma
$-SPTs in the case $\Gamma
=\mathbb Z_{n/2}^{\eta^2}\rtimes\mathbb Z_2^{\eta U}$ and we denote the level as $(k_1,k_2)$ where $k_1, k_2\in \Z_2$. To discuss how $U$
acts on these SPTs, we start with the example with $n=4$ where $\Gamma=\Z^{\eta^2}_2\times \Z^{\eta U}_2$. The $\Gamma$-SPT phase can be classified by the topological response action \cite{PhysRevB.91.035134,PhysRevLett.114.031601}:
\[
\exp\left(\frac{i\pi}{2} k_1 \int_{M_4}ABdB +\frac{i\pi}{2} k_2 \int_{M_4}BAdA\right), 
\]
where $A,B$ is $\Z_2$ background gauge field of $\Z^{\eta^2}_2$ and $\Z^{\eta U}_2$ respective  and $M_4$ is a four dimensional spacetime manifold. Due to eq.\eqref{eq:H trans2},
$U$ acts on $\Z_2$ background gauge field as
\[
U: A\to -A, B\to A+B.
\]
The topological action is mapped to 
\[
\exp\left(\frac{i\pi}{2} k_1 \int_{M_4}A(A+B)d(A+B) +\frac{i\pi}{2} k_2 \int_{M_4}(A+B)AdA\right)= \exp\left(\frac{i\pi}{2} k_1 \int_{M_4}ABdB +\frac{i\pi}{2} (k_1+k_2) \int_{M_4}BAdA\right).
\]
Hence the SPT with level $(0,0)$ and $(0,1)$ is invariant under $U$ but the level $(1,0)$ and $(1,1)$ is connected by $U$. This implies there are three distinct conjugacy classes.

The anomalous symmetries on the boundary theory of the above SPT phases can be further realized on the triangle lattice as \cite{yoshida2016topological,PhysRevB.106.224420}:
\begin{enumerate}
    \item When $k_1=0,k_2=1$, $U_e(\eta^2)=\prod_{j}\sigma^x_j$, $U_e(\eta U)=\prod_{(i,j,k)\in\bigtriangleup/\bigtriangledown}\text{CCZ}^{\sigma}_{i,j,k}$.  Physically, the response action is consistent with the fact that $U_e(\eta U)$ is the entangler of two spatial dimensional $\Z^{\eta^2}_2$-SPTs.
    \item When $k_1=1,k_2=0$, $U_e(\eta^2)=\prod_{(i,j,k)\in\bigtriangleup/\bigtriangledown}\text{CCZ}^{\sigma}_{i,j,k}$, $U_e(\eta U)=\prod_{j}\sigma^x_j$. Now  $U_e(\eta^2)$ is entangler of two spatial dimensional $\Z^{\eta U}_2$-SPTs, which is consistent with the response action.
\end{enumerate}
Here we assign a qubit on each vertex and $\text{CCZ}^{\sigma}_{i,j,k}$  is three qubits controlled-Z gate. Then after $U$ transformation, we have the following results.  
\begin{enumerate}
    \item When $k_1=0, k_2=1$, $U'_e(\eta^2)=U_e(\eta^{-2})=\prod_{j}\sigma^x_j$, $U'_e(\eta U)=U_e(\eta^{-1} U)=\prod_j \sigma^x_j\prod_{(i,j,k)\in\bigtriangleup/\bigtriangledown}\text{CCZ}^{\sigma}_{i,j,k}$. Based on the response theory, this symmetry, which is obtained by acting $U$, is the anomalous symmetry on the boundary of SPT with level $(0,1)$.
    \item When $k_1=1,k_2=0$, $U'_e(\eta^2)=U_e(\eta^{-2})=\prod_{(i,j,k)\in\bigtriangleup/\bigtriangledown}\text{CCZ}^{\sigma}_{i,j,k}$, $U'_e(\eta U)=U_e(\eta^{-1} U)=\prod_j\sigma^x_j\prod_{(i,j,k)\in\bigtriangleup/\bigtriangledown}\text{CCZ}^{\sigma}_{i,j,k}$. Based on the response theory, this symmetry, which is obtained by acting $U$, is the anomalous symmetry on the boundary of SPT with level $(1,1)$. This implies $U'_e(\eta^2)$ and $U'_e(\eta U)$ is the entangler of $(2+1)D$ $\Z^{\eta U}_2$-SPT and  $\Z^{\eta^2}_2$-SPT respective.
\end{enumerate}

Then we generalize the above lattice realization to general $n=0$ (mod 4). Here we realize anomalous symmetries on the boundary theory where  we assign a qubit and a $\mathbb Z_{n/2}$ qudit on each vertex of the triangle lattice.
\begin{enumerate}
    \item When $k_1=0,k_2=1$, $U_e(\eta^2)=\prod_j X_j\sigma^x_j$ and $U_e(\eta U)=\prod_j C_j\prod_{(i,j,k)\in\bigtriangleup/\bigtriangledown}\text{CCZ}^{\sigma}_{i,j,k}$. $U_e(\eta U)$ is $\Z_2$-entangler of two spatial dimensional $\Z^{\eta^2}_{n/2}$-SPTs. 
    \item When $k_1=1,k_2=0$,  $U_e(\eta^2)=\prod_j X_j \prod_{(i,j,k)\in\bigtriangleup/\bigtriangledown}\text{CCZ}^{\sigma}_{i,j,k}$ and $U_e(\eta U)=\prod_j C_j\sigma^x_j$. $U_e(\eta^2)$ is $\Z_2$-entangler of two spatial dimensional $\Z^{\eta U}_2$-SPTs.
\end{enumerate}
 The above result is because the anomaly index of this construction all comes from qubit $\sigma$ degree of freedom, which is the same as the case $n=2$. Moreover, this also implies that the result after $U$ transformation should be the same as the case with $n=2$. In summary, combing with case where $\Gamma=\Z_n^{\eta}$, there are $1+3=4$ different $\Z^U_2$-SSB phases when $n=0$ (mod $4$).

\end{document}